\def\A             {\ensuremath{\mathfrak A}}
\def\Ad            {{\rm Ad}}
\def\alg           {algebra}

\def\bc            {boundary condition}
\def\be            {\begin{equation}}
\def\bearl         {\begin{array}{l}}
\def\bearll        {\begin{array}{ll}}
\newcommand\bG[1]  {{\mathscr B}^\G_{#1}}
\newcommand\BG[1]  {B^\G_{#1}}
\newcommand\boxx[2]{{\fbox{\scriptsize$\!\!\bearl~\\[-1.3em]\kg\eq#2\\
                    \La\eq#1\\[-1.4em]~\eear\!\!\!$}}}
\newcommand\bPF[1] {{\mathscr B}^\PF_{#1}}
\newcommand\bQ[1]  {{\mathscr B}^\Q_{#1}}
\newcommand\BQ[1]  {B^\Q_{#1}}
\newcommand\bZ[1]  {{\mathscr B}^\SUtwo_{#1}}
\def\bzl           {\tilde{\mathscr B}^\SUtwo_{\Lax}} 
 
\def\brsp          {\ensuremath{\mathcal B}}
\def\calh          {\ensuremath{\mathcal H}}
\def\cc            {weight-conjugacy class}   
\def\cdo           {\,{\cdot}\,}
\def\cft           {conformal field theory}
\def\Cft           {Conformal field theory}
\def\cfts          {conformal field theories}
\newcommand\cgc[5] {c_{#2,#3}^{#4 \prec #5;\,#1}}

\newcommand\cgcs[5]{c_{#2,#3}^{#4 \prec #5;\,#1\,^{\scriptstyle*}}}
\def\Chi           {{\mathcal X}}

\def\CLala         {C^{}_{\La,\la}}
\def\CLalap        {C^{}_{\La,\la^+_{\phantom;}}}
\def\CmLala        {C^{\scriptscriptstyle(-)}_{\La,\la}}
\def\cocon         {coset construction}
\def\como          {coset model}
\def\complex       {{\ensuremath{\mathbbm C}}}
\def\con           {conformal }
\def\corfu         {correlation function}
\def\CtLala        {\widetilde C^{}_{\La,\la}}
\def\Dpi           {\ensuremath{D_{\pi/2}}}
\def\dsty          {\displaystyle}

\def\dg            {{\rm d}g\,}
\def\du            {{\rm d}u\,}
\def\dv            {{\rm d}v\,}
\def\ee            {\end{equation}}
\def\eE            {{\rm e}}
\def\eear          {\end{array}}
\def\embed         {\hookrightarrow}
\def\Embed         {\,{\embed}\,}

\def\eq            {\,{=}\,}
\newcommand\erf[1] {(\ref{#1})}
\def\expits        {{\eE^{\ii t \sigma_3}}}
\def\F             {{\mathscr F}}

\def\FG            {\ensuremath{\F(\G)}}
\def\FGtH          {\ensuremath{\F(\G{\times}\HO)}}
\def\FH            {\ensuremath{\F(\HO)}}
\def\FQ            {\ensuremath{\F(\Q)}}
\def\findim        {fi\-ni\-te-di\-men\-si\-o\-nal}
\def\FJj           {F_\Jj}
\def\FM            {\ensuremath{\F(\target)}}
\newcommand\Frac[2]{\mbox{\large$\frac{#1}{#2}$}}
\def\frob          {Fro\-be\-ni\-us algebra}

\def\fus           {{*}}
\def\fusb          {{\overline*}}
\def\g             {\ensuremath{\widehat{\liefont g}}}
\def\G             {\ensuremath{{\Liefont G}}}
\def\gbar          {\ensuremath{{\liefont g}}}
\def\gbarR         {\ensuremath{\gbar_\reals}}
\def\gh            {\ensuremath{{\gbar}{/}{\hbar}}}
\def\GH            {\ensuremath{{\G}{/}{\H}}}
\def\GHad          {\ensuremath{{\G}{/}{\Ad(\H)}}}
\def\ghk           {\ensuremath{(\gh)_\kg}}
\def\gJ            {\ensuremath{g_{(\J)}^{}}}
\def\GHlr          {\ensuremath{(\GtH){/}(\HlHr)}}
\def\GHLR          {\ensuremath{\frac{\G{\times}\HO}{\HlHr}}}
\def\gL            {{g_\La^{\phantom:}}}
\def\go            {\gbar_\circ}
\def\GtH           {\ensuremath{\G{\times}\HO}}
\def\gth           {\ensuremath{\gbar{\times}\hs}}
\def\gv            {\ensuremath{{\levelfont g}^\vee}}
\def\h             {\ensuremath{\widehat{\liefont h}}}
\def\H             {\ensuremath{{\Liefont H}}}
\def\Haa           {\ensuremath{{\mathcal H}_{\rm A}^{aa}}}
\def\Haahor        {\ensuremath{\overline{\mathcal H}{}_{\rm A}^{aa}}}
\def\Hab           {\ensuremath{{\mathcal H}_{\rm A}^{ab}}}

\def\hbarR         {\ensuremath{\hbar_\reals}}
\def\hj            {\ensuremath{h_{(\j)}}}
\def\hjm           {\ensuremath{h_{(\j)}^{-1}}}
\def\Hl            {\ensuremath{\H_l^{}}}
\def\HlHr          {\ensuremath{\Hl{\times}\Hr}}

\def\HO            {\H}
\def\Hr            {\ensuremath{\H_r^{}}}
\def\hs            {\ensuremath{\overline\hbar}}
\newcommand\hsp[1] {\mbox{\hspace{#1 em}}}
\def\Htor          {\ensuremath{{\mathcal H}_{\rm T}}}
\def\Htorhor       {\ensuremath{\overline{\mathcal H}_{\rm T}}}
\def\hv            {\ensuremath{{\levelfont h}^\vee}}
\def\hw            {highest weight}
\def\hy            {$\mbox{-\hspace{-.66 mm}-}$}
\def\Id            {\ensuremath{{\mathscr J}_{\gh}}} 
\def\Ido           {\ensuremath{{\mathscr J}_{\gbar/\go}}} 
\newcommand\iG[1]  {{\mathcal I}^\G_{#1}}
\newcommand\IG[1]  {I^\G_{#1}}
\def\igh           {{\rm i}_{\hbar\Embed\gbar}}
\newcommand\iGH[2] {{\mathcal I}^{\G\times\HO}_{#1,#2}}
\def\ii            {{\rm i}}
\def\ij            {{\ell}}
\def\ik            {{\ell'}}
\def\iN            {\,{\in}\,}

\def\intG          {\int_{\!\G}\!}
\def\intHH         {\int_{\!\HO\times\HO}\!\!}
\def\intHHn        {\frac1{\VH^2_{}}\intHH}
\newcommand\IQ[1]  {I^\Q_{#1}}
\newcommand\iQ[1]  {{\mathscr I}^\Q_{#1}}

\def\j             {{\ensuremath{\jmath}}}
\def\J             {{\ensuremath{J}}}
\def\Jg            {\ensuremath{{\mathscr J}_{\gbar}}} 
\def\JgJh          {\ensuremath{\Jg{\times}\Jh}} 
\def\Jh            {\ensuremath{{\mathscr J}_{\hbar}}} 
\def\Jj            {{\ensuremath{(\J,\j)}}}
\def\kg            {{\ensuremath{\levelfont k}}}
\def\kh            {\ensuremath{{\levelfont k}'}}
\def\la            {{\lambda}}
\def\La            {{\Lambda}}
\newcommand\labl[1]{\label{#1}\ee}
\def\lala          {{[\Lax,\lax]}}
\def\lalab         {{(\La,\la)}}
\def\lao           {{\lambda}}
\def\Lao           {{\Lambda}}
\def\lalao         {{[\La,\la]}}
\def\lalaoprime    {{[\La',\la']}}
\def\lalaprime     {{[\Lax',\lax']}}
\def\lalapsi       {{[\Lax,\lax;\psi]}}
\def\lap           {{\lambda^{\!+}_{\phantom:}}}
\def\Lap           {{\Lambda^{\!+}_{\phantom:}}}
\def\lax           {{\widehat\la}}
\def\Lax           {{\widehat\La}}
\def\levelfont     {\rm }
\def\lie           {Lie algebra}
\def\Lie           {Lie group}
\def\liefont       {\mathfrak }
\def\Liefont       {\rm }
\def\lj            {\ensuremath{\la_{(\j)}}}
\def\LJ            {\ensuremath{\La_{(\J)}}}
\def\lr            {{\rm l|r}}
\def\Mapsto        {\,{\mapsto}\,}
\def\multlala      {{\ensuremath{b_{\La,\la}}}}
\def\multlalaz     {{\ensuremath{b_{\La,\la}^{\,2}}}}
\def\One           {\mbox{\small $1\!\!$}1}
\def\onedim        {one-di\-men\-sio\-nal}
\def\oti           {\,{\otimes}\,}
\def\parfu         {partition function}
\def\PF            {{\rm PF}}
\def\piad          {\ensuremath{\pi_\Ad}}
\def\pilr          {\ensuremath{\pi_\lr}}
\def\pilrs         {\ensuremath{\pi_\lr^\star}}
\def\pl            {\,{+}\,}
\def\popo          {Po\-in\-ca\-r\'e polynomial}
\def\q             {quantum }
\def\Q             {\ensuremath{{\Liefont Q}}}
\def\QPF           {\ensuremath{{\Liefont Q}_\PF}}
\newcommand\query[1]{\ifnum\draftcontrol=1
                   \marginpar{{~}\\[-4em]\small label:\\``#1"\\[1em]~}\fi}
\def\qzn           {quantization}
\def\reals         {{\ensuremath{\mathbb R}}}
\def\rep           {repre\-sen\-ta\-ti\-on}
\def\rhs           {right hand side}
\def\sicu          {simple current}
\def\ssj           {{\scriptstyle\,(\ij)}}
\def\sssj          {{\scriptscriptstyle\,(\ij)}}
\def\stablala      {\ensuremath{{\mathscr J}_\lala}} 
\def\suthree       {\ensuremath{{\liefont{su}}(3)}}
\def\SUthree       {\ensuremath{{\Liefont SU}(3)}}
\def\sutwo         {\ensuremath{{\liefont{su}}(2)}}
\def\SUtwo         {{\ensuremath{{\Liefont SU}(2)}}}
\def\T             {\ensuremath{{\Liefont T}}}
\def\target        {\ensuremath{\mathcal M}} 
\def\targeta       {\ensuremath{\target_a}} 
\newcommand\tbPF[1]{\tilde{\mathscr B}^\PF_{#1}}
\def\tcs           {tensor categories}
\def\tcsa          {{\rm t}}  
\def\til           {\tilde }
\newcommand\till[1]{\tilde{#1}_{\ij}}
\newcommand\tilk[1]{\tilde{#1}_{\ik}}
\def\twodim        {two-di\-men\-sio\-nal}
\def\Uone          {\ensuremath{{\Liefont U}(1)}}
\def\uone          {\ensuremath{{\liefont u}(1)}}
\def\uoneh         {\ensuremath{\widehat{\liefont u}(1)}}

\def\VG            {{\ensuremath{|\G|}}}
\def\VH            {{\ensuremath{|\HO|}}}
\newcommand\void[1]{}
\def\VT            {{\ensuremath{|\T|}}}
\def\WZW           {Wess\hy Zu\-mi\-no\hy Wit\-ten }
\def\wzwm          {WZW model}
\newcommand\xyaxisulevel[1]{\begin{picture}(0,0)
                   \put(10,100)  {\scriptsize\fbox{$\kg\eq #1$}}
                   \put(-6.5,-7) {\tiny$-\frac\pi2$}
                   \put(25,-7)   {\tiny$-\frac\pi4$}
                   \put(60.5,-7) {\tiny$0$}
                   \put(90,-7)   {\tiny$\frac\pi4$}
                   \put(121,-7)  {\tiny$\frac\pi2$} \end{picture}}
\def\zet           {\ensuremath{\mathbb Z}}
\def\zetpo         {\ensuremath{{\mathbb Z}_{\ge0}}}
\def\Zad           {\ensuremath{{\Liefont Z}(\G{,}\H)}}
\def\ZG            {{\ensuremath{{\Liefont Z}(\G)}}}
\def\ZH            {{\ensuremath{{\Liefont Z}(\H)}}}
\def\Zlr           {\ensuremath{{\Liefont Z}_\lr}}

\catcode`\@=11

\newif\if@fewtab\@fewtabtrue
{\count255=\time\divide\count255 by 60 \xdef\hourmin{\number\count255}
\multiply\count255 by-60\advance\count255 by\time
\xdef\hourmin{\hourmin:\ifnum\count255<10 0\fi\the\count255}}
\def\ps@draft{\let\@mkboth\@gobbletwo \def\@oddhead{}
    \def\@oddfoot{\hbox to 7 cm{\tiny \versionno
       \hfil}\hskip -7cm\hfil\rm\thepage \hfil {\tiny\draftdate}}
    \def\@evenhead{}\let\@evenfoot\@oddfoot}
\def\draftdate{\number\month/\number\day/\number\year\ \ \ \hourmin }
 \global\def\draftcontrol{0}
 
\def\draftcite#1{\ifnum\draftcontrol=1#1\else{}\fi}
\def\@lbibitem[#1]#2{\item{}\hskip -3\hbox to 2cm
{\hfil$\scriptstyle\draftcite{#2}$}\hskip
1cm[\@biblabel{#1}]\if@filesw {\def\protect##1{\string ##1\space}\immediate
      \write\@auxout{\string\bibcite{#2}{#1}}}\fi\ignorespaces} 
\def\@bibitem#1{\item\hskip -3cm \hbox to 2cm
{\hfil {\footnotesize\draftcite{#1}}}\hskip 1cm
\if@filesw \immediate\write\@auxout
       {\string\bibcite{#1}{\the\value{\@listctr}}}\fi\ignorespaces}

\def\citen#1{\if@filesw \immediate\write \@auxout {\string\citation{#1}}\fi%
\@tempcntb\m@ne \let\@h@ld\relax \def\@citea{}%
\@for \@citeb:=#1\do {\@ifundefined {b@\@citeb}%
    {\@h@ld\@citea\@tempcntb\m@ne{\bf ?}%
    \@warning {Citation `\@citeb ' on page \thepage \space undefined}}%
    {\@tempcnta\@tempcntb \advance\@tempcnta\@ne
    \setbox\z@\hbox\bgroup\ifcat0\csname b@\@citeb \endcsname \relax
    \egroup \@tempcntb\number\csname b@\@citeb \endcsname \relax
    \else \egroup \@tempcntb\m@ne \fi \ifnum\@tempcnta=\@tempcntb
    \ifx\@h@ld\relax \edef \@h@ld{\@citea\csname b@\@citeb\endcsname}%
    \else \edef\@h@ld{\hbox{--}\penalty\@highpenalty
    \csname b@\@citeb\endcsname}\fi
    \else \@h@ld\@citea\csname b@\@citeb \endcsname \let\@h@ld\relax \fi}%
\def\@citea{,\penalty\@highpenalty\hskip.13em plus.13em minus.13em}}\@h@ld}
\def\@citex[#1]#2{\@cite{\citen{#2}}{#1}}%
\def\@cite#1#2{\leavevmode\unskip\ifnum\lastpenalty=\z@\penalty\@highpenalty\fi%
  \ [{\multiply\@highpenalty 3 #1%
  \if@tempswa,\penalty\@highpenalty\ #2\fi}]}   %
\makeatother 

\catcode`\@=12

\documentclass[12pt]{article} 
\usepackage{latexsym,amsmath,amssymb,amsfonts,bbm,epsf,color,colordvi}
\usepackage[dvips]{graphics} 
\usepackage[mathscr]{eucal}
\usepackage{rotating}
\numberwithin{equation}{section}
\def\hbar          {\ensuremath{{\liefont h}}}
\begin{document}

\begin{flushright}  {~} \\[-12mm] {\sf hep-th/0505117}\\[1mm]
{\sf May 2005} \end{flushright}
\begin{center} \vskip 26mm
{\Large\bf ON THE GEOMETRY OF COSET BRANES}\\[24mm]
{\large J\"urgen Fuchs \ \ \ and \ \ \ Albrecht Wurtz}\\[12mm]
Karlstads Universitet\\[2mm] Universitetsgatan 5 \\[2mm] 
S\,--\,651\,88\, Karlstad
\end{center}
\vskip 28mm

\begin{quote}{\bf Abstract}\\[1mm]
Coset models and their symmetry preserving branes are studied from a
representation theoretic perspective, relating e.g.\ the horizontal branching 
spaces to a truncation of the space of bulk fields, and accounting for
field identification. This allows us to describe the fuzzy geometry 
of the branes at finite level.
\end{quote} \vfill \newpage

\section{Introduction}

There are two conceptually rather different approaches to \twodim\ \cft\ --
\rep\ theoretic on the one hand, and geometric on the other. In the former 
approach, at the chiral level the basic structures are the chiral symmetry 
algebra -- a conformal vertex algebra \A\ -- and its \rep s (see e.g.\ 
\cite{BAki,FRbe,huan24}), while the full non-chiral theory is obtained by 
combining the chiral information with algebraic structures in the \rep\ 
category of the chiral algebra \cite{fuRs,fuRs10}.
In the geometric description, one deals instead with a sigma model on a
suitable target space, which can be studied by Lagrangian field theory
methods; see, for instance, \cite{gawe13,gawe16}. Clearly, on either side a 
lot can be learned by understanding how the two descriptions are related.

For closed world sheets, the relation between the two approaches is intimately 
linked to the interpretation of the state space \Htor\ for the torus  -- 
the space of bulk fields, or of closed string states. 
In the \alg ic framework, \Htor\ is described as a \rep\ space of the direct 
sum $\A\,{\oplus}\,\A$ of two copies\,%
  \footnote{~or, for heterotic theories, of different left and right chiral
   algebras.} 
of the chiral algebra; the character of 
this \rep\ is the torus partition function $Z$. In the geometric setting one
would like to interpret a subspace \Htorhor\ of \Htor\ as a space \FM\ of 
functions on the target space manifold \target, implying in particular that 
\Htorhor\ carries the structure of a commutative algebra.

Owing to the truncation to a subspace \Htorhor, a relation between the geometric
and \rep\ theoretic description is most immediately established for models which
come in families depending on some parameter, in a `semiclassical' limit for 
that parameter -- e.g.\ when the level \kg\ of a WZW model gets large. 
(There are various inequivalent ways of performing such limits; for a discussion
see e.g.\ \cite{olrs,afmo2,haob3,fuSc3,roWe} and chapter 16.3 of \cite{FRbe2}.)
In this case a potential description of \Htorhor\ is as the subspace of those 
states in \Htor\ whose conformal weight tends to zero in the limit. On the other 
hand, at finite values of the parameter an algebra structure on \Htorhor\ 
will typically be non-commutativealgebra. Accordingly, \Htorhor\ should rather 
be understood as a space of functions on a non-commutative manifold, which 
may be regarded as a quantized version of \target\ \cite{frgA,sewi8,szab3}.

      \medskip

When world sheets with non-empty boundary are admitted, one must in addition 
account for the state spaces \Hab\ for the annulus with specified conformal 
boundary conditions $a$ and $b$ -- the spaces of boundary fields, or of open 
string states.  
In particular, to each boundary condition $a$ there is associated the space 
\Haa.  Each \Hab\ is a \rep\ space of a single copy of the chiral algebra \A;
its character is the annulus amplitude ${\rm A}^{ab}_{}$. Geometrically, 
an elementary boundary condition $a$ is, in the simplest case, described as a 
submanifold \targeta\ of \target\ \cite{polc8,gawe18}. In the string theory 
context, \targeta\ plays the role of a D-brane -- the submanifold on which 
open strings can end. In a semiclassical limit (if it exists), a suitable 
subspace \Haahor\ of \Haa\ may be interpreted as the commutative algebra 
of functions on the brane world volume (and in the same limit, open strings 
connecting different branes disappear). Again, at finite values of the relevant 
parameter, and taking into account the background $B$-field, \Haahor\ carries 
the structure of a non-commutative algebra and should be interpreted as a space 
of functions on a non-commutative quantization of \targeta\
\cite{scho5,gapl,alrs3,paSt,afqs}. (Accordingly, the low-energy dynamics
of the gauge fields on the brane is described by a non-commutative Yang\hy Mills
theory. Here we neither address the algebraic structure of \Haahor\ nor brane
dynamics, though.)

      \medskip

The geometric intepretation of \bc s is quite well understood for (unitary)
\WZW models, for which the target space is a compact reductive Lie group \G. 
In particular, the symmetry preserving branes, labeled by the integrable 
\rep s of the relevant affine Lie algebra \g\ at level \kg,
are conjugacy classes of \G, while a specific
class of symmetry breaking branes is given by `twined'  conjugacy classes;
see e.g.\ \cite{alsc2,gawe14,fffs,staN5,quel2,mowe,gawe18}. 
Since many more rational \cft\ models can be obtained from WZW models via
the coset construction, it is natural to try to extend this analysis to
coset models, for which the target is a corresponding quotient \Q\ of compact 
reductive Lie groups. Indeed, a lot of effort has been devoted to the study 
of \bc s in coset models, a partial list of references being
\cite{gawe16,staN3,maMS,elsa,isHi,frsc3,krwz,noza,gagR2}. However, it is fair 
to say that the situation is less satisfactory than in the WZW case, and some 
interesting questions are still open. It is the purpose of the present paper 
to collect, with the help of \rep\ theoretic tools, further knowledge about 
the geometry of branes in coset models.

Among the outcome, what seems to us most relevant is a
characterization of the group of identification simple currents
in terms of the functions \FQ\ on the target space (see formula \erf{gJ=hj})
and the implementation of field identification in the geometric description 
of the branes (see the discussion around \erf{iQ}). To a certain extent our 
discussion is in semiclassical (large level) spirit, but primarily we are 
interested in what happens at finite values of the level. After all, in any 
WZW or coset model, the level does have a definite finite value.

\medskip

The paper is organized as follows.
In section 2 we collect information about algebraic aspects of \como s
that is needed. In section 3 the geometry of \como s is discussed, leading
in particular to a description of \Htorhor\ in terms of horizontal branching
spaces (formula \erf{FBB}). The study of coset branes starts in section 4
with considerations related to the semiclassical limit of large level,
while section 5 contains a detailed description of the branes at finite
level. As an application we clarify in section 6 the geometry of 
parafermion branes at finite level. 
In an appendix, we comment on general aspects of the large level limit.


\section{Coset models}

Algebraically, a coset model is determined by an
embedding $\h\Embed\g$, where \g\ and \h\ are 
direct sums of untwisted affine \lie s and Heisenberg \alg s
(with identified centers), together with the choice of a level \kg\
of \g-\rep s, that is, a level for each of the affine ideals of \g\ as well 
as a corresponding integer specifying the rational extension of the vertex 
algebra associated to each \uoneh\ ideal of \g. Here we only 
consider unitary theories, thus (each component of) \kg\ is a 
positive integer. The chiral algebra \A\ of the \como\ is the commutant 
of the vertex \alg\ associated to \h\ in the vertex \alg\ associated to \g.

Further, an allowed embedding $\h\Embed\g$ must come from a corresponding 
embedding $\hbar\Embed\gbar$ of the respective horizontal subalgebras (which 
are \findim\ complex reductive \lie s), from which it is obtained via the loop 
construction. Given the level \kg\ of \g, the \h-\rep s are at level $\kh\eq
\igh\kg$, with $\igh$ the Dynkin index of the embedding $\hbar\Embed\gbar$.

The coset model obtained this way is often denoted by \GH, where \G\
and \H\ are Lie groups associated to \g\ and \h\ (see
the next section). But the relevant target \Q\ is not the standard space of
cosets of \G\ by the right (or left) action of \H, and accordingly we prefer 
the notation \gh\ or, when we also want to indicate the level of the relevant 
\g-\rep s, by \ghk. This notation fits well with the fact that, as a rational 
\cft, the \gh\ \como\ can be analyzed completely through the \rep\ theory 
of \g\ and of \h, and thereby through the corresponding WZW models,
to which we will refer as the \gbar- (or $\gbar_\kg$-) and \hbar-\wzwm s,
respectively. Indeed, the \gh\ model can be realized as a suitable extension
of the tensor product \gth\ of the \gbar-\wzwm\ and a putative\,%
  \footnote{~One way to think of \hs\ is as a combination of a \hbar-\wzwm\ at 
  negative levels and a ghost system \cite{hwrh}.}
theory \hs. 
The latter theory \hs\ is obtained from the \hbar-\wzwm\ by a specific 
modification of the \rep\ category of its chiral algebra which, basically, 
amounts to taking the complex conjugate of all chiral data, hence in particular 
of the modular group \rep\ (for details and references see \cite{ffrs2}).

For the purposes of this paper we restrict our attention to those \como s for
which the labelling of fields is related to the one for the \gbar\ and \hbar\ 
theories by means of \sicu s \cite{scya}.
(Otherwise the \como\ is called a maverick coset.\,%
  \footnote{~See again \cite{ffrs2} for more information and references.
  The maverick cosets include in particular all conformal embeddings 
  $\h\Embed\g$, for which the coset theory has zero Virasoro central charge
  and hence is trivial.}%
)
Then the primary fields of the \gh\ model are labeled by
equivalence classes $\lalapsi$, where $\Lax$ 
is an integrable highest \g-weight of level \kg\ and $\lax$ an integrable 
highest \h-weight of level \kh, while $\psi$ is a certain degeneracy label
\cite{mose4,gepn8,scya5,fusS4,fuRs9}. These classes are orbits of the action 
of certain \sicu s on (the labels of) primary fields by the fusion product 
(to be denoted by the symbol $\fus$), i.e.~the equivalence relation reads 
  \be
  (\Lax,\lax;\psi) \,\sim\,
  (\J\fus\Lax,\j\fus\lax;p_\Jj(\psi)) \qquad {\rm for}\quad \Jj\iN\Id 
  \labl{field-id}
with
  \be
  \Id \,\subseteq\, \Jg\,{\times}\,\Jh 
  \labl{Id}
a specific subgroup of $\Jg{\times}\Jh$, where \Jg\ and \Jh\ are the groups 
of \sicu s of the \gbar- and \hs-theories, respectively.\,%
  \footnote{~The exceptional simple current of $\widehat E_8$ at level two
  is excluded here.}
This subgroup \Id\ is called the {\em identification group\/} of the \gh\ 
model, and its elements are referred to as identification currents. 
Also, $p_\Jj$ in \erf{field-id} is a permutation
of the degeneracy labels (compare formula (A.9) in \cite{fuSc11}).
In the sequel, we will for simplicity assume that the `field identification' 
\erf{field-id} does not have fixed points, i.e.\ that all stabilizer subgroups
  \be
  \stablala := \{ \Jj\iN\Id \,|\, \J\fus\Lax\eq\Lax~{\rm and}~
  \j\fus\lax\eq\lax \}  
  \labl{stablala}
of \Id\ are trivial, so that the degeneracy labels $\psi$ are absent.
Let us remark, however, that we do expect that some of our considerations can be
extended to \como s with field identification fixed points, though this is
definitely beyond the scope of the present paper. On the other hand, it should
not be expected that this framework allows one to understand the behavior of 
maverick cosets, which seem to be a low level phenomenon.

Further, among the classes $\lala$ only those appear as labels of primary 
fields of the \gh\ model for which the branching space
$\brsp_{\Lax,\lax}\eq{\rm Hom}_{\h}(\calh_\lax,\calh_\Lax)$ 
is non-zero, i.e.\ for which the irreducible \h-module $\calh_\lax$ with 
\hw\ $\lax$ occurs in the decomposition
  \be
  \calh_\Lax \eq \bigoplus_\lax^{} \brsp_{\Lax,\lax} \oti \calh_\lax
  \labl{HBHx}
of the irreducible \g-module $\calh_\Lax$ with \hw\ $\Lax$:
  \be
  \lala \mbox{~~allowed label~~} \Leftrightarrow\;\brsp_{\Lax,\lax}\ne\{0\}
  \,. \labl{selrule}
In the absence of field identification fixed points, the branching spaces 
$\brsp_{\Lax,\lax}$ constitute the irreducible modules of the coset chiral 
algebra, and the branching functions, i.e.\ the generating functions for the 
dimensions of the subspaces with definite conformal weight
of branching spaces, are the irreducible characters of the coset theory.
Also, branching spaces related by the field identification \erf{field-id} are
isomorphic as modules of the coset chiral algebra, 
$\brsp_{\Lax,\lax}\,{\cong}\,\brsp_{\J\fus\Lax,\j\fus\lax}$ for $\Jj\iN\Id$.

Just like the field identification, the selection rule \erf{selrule} can be 
understood in terms of the group \Id, too: precisely those pairs $(\Lax,\lax)$ 
are allowed whose monodromy charge \cite{scya} with respect to all 
identification \sicu s $\Jj\iN\Id$ vanishes \cite{scya3}.
It is also well known that for non-maverick cosets the selection rules are 
purely `group-theoretical' in the terminology of \cite{scya5}, i.e.\ that they
are equivalent to imposing compatibility among the \cc es\,%
  \footnote{~By the {\em \cc\/} of a \gbar-weight $\La$ we mean the class of
  $\La$ in the weight lattice modulo the root lattice of \gbar. We avoid the
  more common term conjugacy class in order that no confusion with
  the conjugacy classes $C$ in the Lie group \G\ associated (see below) to
  \gbar\ can arise.}
of the respective horizontal projections $\La$ and $\la$ of the weights $\Lax$ 
and $\lax$.  The latter description allows one in particular to determine the 
identification group directly from the relevant embedding $\hbar\Embed\gbar$ 
of horizontal subalgebras (compare e.g.\ \cite{levw,sche3,fuSc}),
though to the best of our knowledge this hasn't been 
investigated in detail for arbitrary (non-maverick) \como s.


\section{Geometry of coset models} \label{secGocm}

In the Lagrangian framework, the \gh\ coset model is obtained by coupling
the WZW model with target space \G\ -- the
connected and simply connected compact \Lie\ whose \lie\ is
the compact real form \gbarR\ of \gbar\ -- to a gauge field taking values in
\hbarR\ \cite{bars,gaku2,kpsy}.  By integrating out the gauge field in the
path integral, this yields a sigma model with target $\target\eq\Q$ given by
the space \GHad\ of orbits of the adjoint action of a subgroup \H\ on \G\
\cite{witt36},
  \be
  \Q = \GHad := \{ [g] \,|\, g\iN\G \} \qquad{\rm with}\qquad
  [g] := \{ hgh^{-1} \,|\, h\iN\H \} \,.
  \labl{GHad}
Note that here we have to think of \H\ not just as a \Lie, but rather as a Lie 
subgroup embedded in \G\ via a concrete embedding $\imath{:}\ \H\Embed\G$, 
which is determined by the \lie\ embedding $\hbar\Embed\gbar$. 
Accordingly, in the sequel group elements $h\iN\H$ will always be regarded 
as elements of \G, i.e.\ $h\,{\equiv}\, \imath(h)\iN\G$.

Also, in \erf{GHad} we take \HO\ to be the connected and simply connected 
compact \Lie\ whose \lie\ is \hbarR. However, the subgroup 
  \be
  \Zad := \ZG \cap \HO
  \ee
of \HO, with \ZG\ the center of \G, acts trivially under the adjoint action, 
so that in \erf{GHad} we could equally well replace \HO\ by the (generically) 
non-simply connected group $\HO/\Zad$. Indeed, in the description of the \gh\ 
model as a gauged \wzwm, the gauge field is a connection in a principal bundle 
with structure group $\HO/\Zad$ rather than \HO.\,%
  \footnote{~To deal with fixed point resolution, also gauge fields in
  non-trivial $\HO/\Zad$ bundles must be included \cite{mose4,hori,schW3}.}
The significance of this observation for the \gh\ model becomes evident when one
notices that the centers \ZG\ and \ZH\ are naturally isomorphic to (the duals 
of) the groups of \cc es of \gbar- and \hbar-weights, respectively. The 
triviality of the action of \Zad\ therefore corresponds to having a definite
relation between the \cc es of the \gbar- and 
\hbar-weights that can appear as labels of primary fields in the \gh\ model. 
Comparison with the description of the selection rules in the previous
section shows that indeed \Zad\ is isomorphic to the identification group,
  \be
  \Zad \,\cong\, \Id \,.
  \labl{ZadcongId} 

An alternative description of the target space is \cite{frsc3}
  \be
  \Q = \GHLR := \{ [g,h] \,|\, g\iN\G,\, h\iN\HO \} \qquad{\rm with}\qquad
  [g,h] := \{ (ugv,uhv) \,|\,  u,v\iN\HO \}\,, 
  \labl{GHlr}
where \Hl\ and \Hr\ are just equal to \HO\ as Lie subgroups of \G, while their 
subscript reminds us that they act from the left and right, respectively, on 
each factor of \GtH. A bijection from \GHad\ to \GHlr\ is given by 
$\varpi{:}\ [g]\Mapsto[g,e]$ with $e$ the unit element of \G\ (and \H), 
with inverse $\,\varpi^{-1}{:}\ [g,h]\Mapsto[gh^{-1}]$ (note that
$[g,h]\eq[gh^{-1},e]$ in \GHlr). 
Again, a sub\-group $\Zlr\,{\subset}\,\HlHr$ acts trivially, and again
this subgroup is isomorphic to the identification group,
  \be
  \Zlr = \{ (u,u^{-1})\,|\, u\iN\Zad \} \,\cong \Zad \,.
  \ee

We will denote by 
  \be \begin{array}{llll}
  \piad: & \G\to\Q & \quad{\rm and}\qquad \pilr: & \GtH\to\Q \\[2pt]
         & \; g \mapsto [g]    &   &  (g,h) \mapsto [g,h]
  \eear\ee
the projections that correspond to descriptions of \Q\ via \erf{GHad} and 
via \erf{GHlr}, respectively.

\paragraph{Functions on \Q.} 

For studying the geometric interpretation of \bc s,
we will have to work with the space \FQ\ of (square integrable \complex-valued) 
functions on \Q, or on certain subsets of \Q. The functions on \Q\ can be 
identified with the \HlHr-invariant functions on \GtH. Accordingly,
the projection $\pilrs f\iN\FQ$ of a function $f\iN\FGtH$ is given by
  \be
  \pilrs f([g,h]) := \intHHn \du\dv f(ugv,uhv) \,,
  \labl{pilrs}
where $\du$ is the (unnormalized\,%
  \footnote{~Since the action of a WZW sigma model, and hence the metric on the
  group manifold \G\ to be used here, involves a factor of \kg, the volume of
  \G\ scales as $\kg^{{\rm dim}(\G)/2}$ with the level.}%
) 
Haar measure on \HO\ and $\VH\,{=}\int_{\!\HO}\!\du$ the volume of \HO.
A function $f\iN\FGtH$ is $\HlHr$-invariant iff $\pilrs f\eq f$. 

The space \FGtH\ is spanned over \complex\ by the functions
  \be
  D^{\La,\la}_{mn,ab}
  := D^\La_{mn} \, D^{\la_{}^{\,\scriptstyle*}}_{ab}
  \labl{DD}
with $\La$ and $\la$ ranging over the dominant integral weights of \G\ and \HO,
respectively, and $D^\La_{mn}(g)$ and $D^\la_{ab}(h)$, with $m,n\iN\{1,2,...\,,
d_\La\}$ and $a,b\iN\{1,2,...\,,d_\la\}$, the entries of the
corresponding \rep\ matrices (the complex conjugation on $D^\la$ is chosen
for later convenience). The functions \erf{DD} actually form a basis of
\FGtH\ and (when choosing orthonormal bases of the \rep\ spaces) satisfy
orthogonality relations which follow from the relation
$\intG\dg D_{mn}^{\La\,\scriptstyle*}\,D_{m'n'}^{\La'}
\eq d_\La^{-1}\VG \delta_{mm'}\delta_{nn'}\delta_{\La\La'}$
for \FG\ together with the analogous relation for \FH.

Let us investigate the behavior of the basis functions \erf{DD} under the 
projection \erf{pilrs}. It turns out that among the functions \erf{DD}, only 
those $D^{\La,\mu}$ give rise to non-zero functions on \Q\ for which
the irreducible \HO-\rep\ $R^\mu$ with highest weight $\mu$ occurs in the
decomposition of the irreducible \G-\rep\ $R^\La$ with highest weight $\La$
as a \HO-\rep, a property that we will indicate by writing $\mu\,{\prec}\,\La$.
All other functions $D^{\Lambda,\lambda}_{mn,ab}$ have vanishing
projection to \FQ. In short, the space \FQ\ is spanned by the functions
$\pilrs D^{\La,\la}_{mn,ab}$ with $\la\,{\prec}\,\La$.
Note that this constitutes a horizontal counterpart of the selection rules
\erf{selrule} of the \como\ \cite{frsc3} -- when studying functions on \Q, 
precisely those pairs $(\La,\la)$ are allowed labels for which 
$\la\,{\prec}\,\La$, i.e.\ those for which the horizontal branching space 
$\brsp_{\La,\la}\eq\complex^\multlala$ that appears in the decomposition
  \be
  \calh_\La \,\cong\, \bigoplus_{\la\prec\La} \brsp_{\La,\la}\oti\calh_\la
  \labl{HBH}
is non-zero.

To analyze this issue in more detail, we formulate the fact that 
the irreducible \G-\rep\ $R^\La$ decomposes as a
direct sum of irreducible \HO-\rep s $R^\mu$ in terms of the \rep\ matrices.
Namely, upon suitable basis choices, for $h\iN\HO$ the \rep\ matrix $D^\La$ 
decomposes into blocks along the diagonal,
  \be
  D^\La(h)
  = \bigoplus_{\la\prec\La}\bigoplus_{\ij=1}^\multlala D^{\la\sssj}(h) \,,
  \labl{DtoD}
where the summation is over all irreducible \HO-\rep s that appear in
the branching of $R^\La$, counting multiplicities, and where the symbol 
$D^{\la\sssj}$ denotes the matrix block within the big matrix $D^\La$ that 
corresponds to the $\ij$th occurence of $R^\la$ in $R^\La$. 
Thus the labels $\lambda\ssj$ enumerate the matrix blocks, while 
$\lambda$ labels (equivalence classes of) irreducible representations.
For a general choice of orthonormal bases in the \rep\ spaces of \G\ and \HO, 
the equality \erf{DtoD} holds up to a similarity transformation. 
The matrix elements are then related by
  \be
  D^\La_{mn}(h) = \sum_\la\sum_{\ij=1}^\multlala \sum_{a,b=1}^{d_\la}
  \cgcs\ij ma\la\La \cgc\ij nb\la\La\, D^\la_{ab}(h)\,.
  \labl{DccD}
The numbers $\cgc\ij ma\mu\La$ appearing here are the coefficients in the 
expansion
  \be
  e^\La_m = \sum_{\mu,\ij,a}\cgc\ij ma\mu\La\, \beta^{\sssj}\oti e^{\mu}_a
  \ee
of the vectors $e^\La_m$ in the chosen basis of $\calh_\La$ as a linear 
combination of vectors $e^{\mu}_a$ in the chosen bases of the irreducible 
\HO-modules $\calh_\mu$ and $\beta^{\sssj}$ in bases of the multiplicity spaces 
$\brsp_{\La,\la}$ (in particular, $\cgc\ij ma\mu\La$ vanishes unless 
$\mu\,{\prec}\,\La$). Being coefficients of a basis transformation between
orthonormal bases, they form unitary matrices.

Using \erf{DccD} together with the representation property and
the orthogonality relations for the \rep\ matrices, we get
  \be \bearll
  \pilrs D^{\Lambda,\lambda}_{mn,ab}(g,h) \!\!\!&\dsty
  = \intHHn \du \dv \!\!\sum_{p,q,\,d,e}\! 
  D^\La_{mp}(u) D^\La_{pq}(g) D^\La_{qn}(v)
  D^\la_{ad}(u)^*_{} D^\la_{de}(h)^*_{} D^\la_{eb}(v)^*_{}
  \\{}\\[-.8em]&\dsty
  = \frac1{d_\la^{\,2}}\, \sum_{\ij,\ik=1}^\multlala 
  \cgcs\ij ma\la\La \cgc\ik nb\la\La\,
  \sum_{p,q=1}^{d_\La} \sum_{c,d=1}^{d_\la} 
  \cgcs\ik qd\la\La \cgc\ij pc\la\La\,
  D^{\Lambda,\lambda}_{pq,cd}(g,h)
  \,.  \eear\labl{pilrsD}
Provided that the \rhs\ of \erf{pilrsD} is non-zero, up to a scalar factor it 
does not depend on the labels
$m,n$ and $a,b$; hence this way each allowed pair $\lalab$ of dominant
integral \G- and \HO-weights gives rise to $\multlalaz$ functions on \Q.

Let us also remark that the calculations simplify when one makes the
following adapted Gelfand-Zetlin type basis choice: First select
arbitrary orthonormal bases in all $\calh_\mu$ and then, for each $\La$, 
a basis of $\calh_\La$ consisting of basis vectors of the $\calh_\mu$,
counting multiplicities according to \erf{HBH}. This corresponds to formula 
\erf{DtoD} holding exactly, not only up to a similarity transformation
(i.e.\ as an equality of matrices rather than just as an equality between
linear transformations). With this adapted basis choice the branching 
coefficients $c^{\mu\prec\La}$ are $\cgc\ij ma\mu\La\eq\delta_{a\,\till m}\,
\delta_{\mu\prec\La}$, and thus we have
  \be
  D^\La_{mn}(h) = D^{\la}_{\till m\till n}(h)\,,
  \labl{branching} 
where the labels $\till m$ and $\till n$ are the row and column labels of
the matrix block $D^{\la\sssj}$ in $D^\La$ that according to \erf{DtoD} 
correspond to the row and column labels $m$ and $n$ of the big matrix.
The result \erf{pilrsD} then reduces to 
  \be 
  \pilrs D^{\La,\la}_{mn,ab}(g,h) = \left\{ \bearll d_\la^{\,-2}
  \sum_{\ij,\ik}\delta_{a\,\till m}\,\delta_{b\,\tilk n}\,
  \sum_{p,q=1}^{d_\La} D^{\La,\la}_{pq,\till p\tilk q} (g,h)
  & {\rm for}~\la\,{\prec}\,\La \,,
  \\{}\\[-.9em] 0 & {\rm else}\,.  \eear\right.
  \ee
In particular, when the branching rule for $\La$ is multiplicity-free, then 
formula \erf{pilrsD} reduces to $
  \pilrs D^{\Lambda,\lambda}_{mn,ab}(g,h)
  \eq \delta_{a\,\til m}\, \delta_{b\,\til n}\, 
  d_\la^{-2} \sum_{p,q} D^{\La,\la}_{pq,\til p\til q}(g,h) $
for $\la\,{\prec}\,\La$.
(This applies e.g.\ to all branching rules in the description of the unitary
Virasoro minimal models as 
$\sutwo{\oplus}\sutwo{/}\sutwo_{\scriptscriptstyle\rm diag}$ \como s, for which
the coefficients $c^{\mu\prec\La}$ are just ordinary 
Clebsch\hy Gor\-dan coef\-fi\-cients.)

\medskip

To summarize, each allowed pair $\lalab$ provides us with $\multlalaz$ 
functions on \Q. Also, from the construction it is apparent that all these
functions are linearly independent. It is then natural to identify the 
$\multlalaz$-dimensional space spanned by these functions with the tensor 
product $\brsp_{\La,\la}^{}\oti\brsp_{\La,\la}^*$ of the branching space 
$\brsp_{\La,\la}\eq\complex^\multlala$ with its dual space. The latter, in 
turn, is naturally isomorphic to the branching space $\brsp_{\Lap,\lap}$
for the dual modules. Thus we arrive at a description of \FQ\ as
  \be
  \FQ \,\cong\, \bigoplus_{\stackrel{\scriptstyle \La,\la}{\la\prec\La}}
  \brsp_{\La,\la}^{} \oti \brsp_{\Lap,\lap} \,.
  \labl{FBB}
This is nothing but the analogue of the the Peter\hy Weyl isomorphism
  \be
  \FG \,\cong\, \bigoplus_\La \calh_\La^{} \oti \calh_\Lap
  \labl{P-W}
for the functions on the target space \G\ of the \gbar-\wzwm.

It is tempting to think of the horizontal branching spaces $\brsp_{\La,\la}^{}$
as subspaces of the branching spaces $\brsp_{\Lax,\lax}$ that appear in the 
branching rules \erf{HBHx} of \g-\rep s, and thereby to identify \FQ\ with the
horizontal subspace \Htorhor\ of the space of bulk fields, 
  \be
  \FQ \,\approx\,  \Htorhor \subset \Htor \,.
  \labl{FQ=H}
Actually, as a consequence of field identification, a few specific states 
present in the $\brsp_{\La,\la}^{}$ can be absent in the $\brsp_{\Lax,\lax}$.
Indeed, it can happen that several different horizontally allowed pairs 
correspond to one and the same state in the coset model.\,\footnote{~%
    For instance, as noted in \cite{bowu}, for the parafermions (see section 6
    below, where also the notation used here is explained in detail), there is
    precisely one such state: at level \kg\ there exist \kg\ allowed fields
    labeled as $(\kg,n)$, while the number of horizontally allowed pairs of
    the form $(\kg,n)$ is $\kg{+}1$.}
However, this appears to be a low-level effect: the number of such missing 
states is small, and it does not depend on the
level, and hence with increasing level this mismatch becomes less and less 
relevant.

\paragraph{Simple current action on \FQ.}
The `horizontal selection rules' just described are the closest analogue 
of the \gh\ selection rules that can reasonably be expected to hold for \FQ. 
It is thus tempting to seek also for a manifestation of the field identification
\erf{field-id} in the space \FQ. Since the fusion rules are (strongly)
level-dependent, there cannot possibly be any nice relation between the 
functions $D^{\La,\la}$ and $D^{\J\fusb\La,\j\fusb\la}$ (by a slight abuse
of notation, by $\J\fusb\La$ we mean the horizontal part of the affine weight
$\J\fus\Lax$). However, there turns out to be another distinctive role of
the identification group.

For \gbar\ and \hbar\ semisimple, an action of the simple currents $\Jj\iN\JgJh$
on functions on \Q\ can be defined as follows. Let us denote by $\tcsa_\La$ the 
element of the Cartan subalgebra $\go$ of \gbar\ that is dual to the weight 
$\La\iN\go^\star$, i.e.\ $\tcsa_\La\eq(\tcsa,\La) \eq\sum_{i,j}\gamma_{i,j}\,
\tcsa^i\La^j$ with $\gamma$ the inverse of the symmetrized 
Cartan matrix. Further, for \J\ a \sicu\ of the \gbar-\wzwm, denote 
by \LJ\ the corresponding \cite{fuge} cominimal fundamental \gbar-weight and set
  \be
  \gJ := \exp\big( 2\pi\ii\, \tcsa_{\LJ} \big) \,\iN\G\,.
  \labl{gJ}
The group elements $\hj \,{:=}\, \exp\big( 2\pi\ii\,(\tcsa',\lj) \big)\iN\G$ 
are defined analogously for the \sicu s of the \hbar-\wzwm. Now
via the fusion product, the set $\Jg\eq\{\J\}$ of \sicu s of the \gbar-\wzwm\ 
is a finite abelian group, and (see e.g.\ section 14.2.3 of \cite{DIms})
is isomorphic as a group to the center \ZG\ of \G, with an isomorphism provided
by the mapping $\J\,{\mapsto}\,\gJ$. Likewise, the mapping $\j\,{\mapsto}\,\hj$ 
furnishes an isomorphism $\{\j\}\eq\Jh\,{\stackrel{\cong~}\to}\,\ZH$.
For an abelian ideal of \gbar\ (and analogously for \hbar), the situation is
a bit simpler. Indeed, then every primary field is a simple current, and the
set of primary fields endowed with the fusion product is an abelian group that
is naturally isomorphic to a subgroup of $\ZG\eq\G$, and we denote the elements
of that subgroup by $\gJ$.

We now define, for $\Jj\iN\JgJh$, an action $\FJj$ on \FQ\ by
  \be
  \big( \FJj f \big) ([g,h]) := f([g\gJ,h\hj]) 
  \ee
for $f\iN\F(\GHlr)$ and $g\iN\G$, $h\iN\H$. This can be rewritten as
  \be
  \big( \FJj f \big) ([g,h]) = f([g\gJ\hjm,h]) \,,
  \ee
thus implying in particular that the subgroup $\{\Jj\iN \JgJh\,|\,\gJ\eq\hj\}$
of \JgJh\ acts trivially. Now it is precisely in this way that the \cc es of the
weights \LJ\ and \lj\ must be related in order for \Jj\ to form an allowed
pair. Moreover, via $\Jj\,{\mapsto}\,(\gJ,\hj)$ this subgroup is 
isomorphic to \Zad, and hence by comparison with \erf{ZadcongId} we see that
  \be
  \{ \Jj \iN \JgJh \,|\, \gJ\eq\hj \} \,\cong\, \Id \,.
  \labl{gJ=hj}
Thus the identification group \Id\ of the \gh\ model
can be characterized as the subgroup of \JgJh\ acting trivially on \FQ.
This is the counterpart of field identification in a geometric analysis
of the \gh\ model in terms of the functions on the target space \Q. Since 
this invariance property is shared (by definition) by the space \Htor\ of bulk
fields, this observation provides
further evidence for the correctness of the identification \erf{FQ=H}.  


\section{Branes at large level} \label{secBall}

The description of coset models given above together with the knowledge about 
D-branes in \wzwm s can be combined to draw conclusions about the geometry of
D-branes of coset models at large level. In this paper we consider 
only such \bc s which preserve the full chiral symmetry of the \como. 
For these boundary conditions,
at large level the branes on $\GtH$ are concentrated on the submanifolds
  \be
  \CLala := \left\{(g,h)\iN \GtH ,|\, g\iN C_\La^\G,h\iN C_\la^\H \right\}
  = C_\La^\G\,{\times}\,C_\la^\H \,\subset \GtH \,.
  \labl{CLala} 
Here
  \be
  C_\La^\G := \{ g'g_\La^{}\, g'{}^{-1} \,|\, g'\iN\G \}
  \labl{C-La}
is the conjugacy class in \G\ of the group element 
  \be  
  g_\La^{} :=  \exp\Big( \Frac{2\pi\ii}{\kg+\gv}\, \tcsa_{\La+\rho} \Big)\,,
  \labl{g-La}
with $\rho$ and $\gv$ the Weyl vector and dual Coxeter number of \gbar,
respectively, and $\tcsa_{\La+\rho}\eq(\tcsa,\La{+}\rho)$. Analogously, 
$C_\la^\H\eq\{h'h_\la^{}\,h'{}^{-1}\,|\,h'\iN\H\}$ is the conjugacy class in \H\
of $h_\la^{}\,{:=}\,\exp\big(\Frac{2\pi\ii}{\kh+\hv}\,\tcsa'_{\la+\rho'}\big)$,
with $\rho'$ and $\hv$ the Weyl vector and dual Coxeter number of \hbar.
Further, the labels $\La$ and $\la$ are (the horizontal parts of) those of 
the primary fields of the \gbar- and \hbar-\wzwm s, i.e.\ dominant integral
weights satisfying $(\La,\theta)\,{\le}\,\kg$ and $(\la,\vartheta)\,{\le}\,\kh$,
with $\theta$ and $\vartheta$ the highest roots of \gbar\ and \hbar, 
respectively.

Let us  remark that instead of \erf{CLala} we may equally well use the sets
  \be
  \CmLala := \left\{(g,h^{-1})\iN \GtH 
  ,|\, g\iN C_\La^\G,h\iN C_\la^\H \right\} \,\subset \GtH \,.
  \ee
to describe the subsets on which the \GtH-branes are concentrated.
Indeed, this just amounts to choosing a different labelling, owing to
  \be
  \CmLala = \CLalap 
  \labl{CLalap}
with $\la^+$ the \hbar-weight charge conjugate to $\la$. The equality 
\erf{CLalap} holds because $h\iN C_\la^\H$ iff $h^{-1}\iN C_{\la^+}^\H$, 
which in turn follows from the fact that the map $\la\,{\mapsto}\,{-}\la^+_{}$
is a Weyl transformation (namely the one corresponding to the longest
element of the Weyl group of \hbar) and that the Weyl vector is
self-conjugate.

In the description \erf{GHlr} of \Q, we can obtain the branes on the target 
space \Q\ of the \como\ as the projections 
  \be
  \pilr(\CLala) = 
  \left\{ [g,h] \,|\, (g,h) \iN \CLala \right\} 
  \labl{pilrC}
of the sets \erf{CLala}.
In the alternative description \erf{GHad}, one deals instead with 
projections of the sets \cite{elsa,gawe16,frsc3,fred'}
  \be 
  \CtLala := \left\{ gh^{-1}\iN\G \,|\ g\iN C_\La^\G,\;h\iN C_\la^\H
  \,{\subset}\,\G \right\} \,\subset\, \G\,.
  \labl{CtLala}
These subsets are $\Ad(\H)$-invariant, i.e.\ satisfy $u\,\CtLala\,u^{-1} \eq 
\CtLala$ for all $u\iN\H$, and hence they trivially project on \Q, i.e.\
  \be
  \piad(\CtLala) = \left\{ [gh^{-1}]\iN\Q \,|\ g\iN C_\La^\G,\;h\iN C_\la^\H
  \right\} \,\subset\,\Q\,.
  \ee

It follows directly from the existence of the bijection $\varpi$ 
defined after \erf{GHlr} that these descriptions of the coset branes at 
large level are equivalent:
  \be
  \piad(\CtLala) = \pilr(\CmLala) \,.
  \labl{49}

\smallskip

Expressed in terms of functions on \Q, the discussion above amounts to
the statement that the shape of a brane is a delta function on the subset
\erf{49}. This description is adequate in the limit of large level, 
whereas at any finite value of the level the shape of the brane is smeared
about this subset. The extent of localization increases with the level;
this will be analyzed quantitatively in the next section.

To see explicitly that in the limit one indeed deals with a delta function
requires the information that in the WZW case the limit yields (a multiple 
of) a delta function $\mathcal D^\G_\La \eq \delta_{C_\gL}^\G/|C_\gL|$ on a 
conjugacy class $C_\gL$ of the group, see formula \erf{WZWB} in the appendix.
The corresponding function on the coset is (including a compensating factor 
of $\VH$, see the discussion after \erf{pDll} below)
  \be
  \mathcal D_{\La,\la}([g,h])
  = \frac1{\VH}\intHH \du\dv \mathcal D^\G_\La(ugv)\,\mathcal D^\HO_\la(uhv)\,.
  \ee
One of the two integrations over \HO\ is trivial, and the other can be
performed with the help of the identity
$\delta_{C_\gL}^\G\!(vuv^{-1}h^{-1}g) \eq \delta_{C_\gL}^\G\!(uh^{-1}g)$,
which for $u,v,h\iN\HO$ and $g\iN\G$ is valid as an equality of functions on \Q.
The result is
  \be
  \mathcal D_{\La,\la}([g,h])
  = |C_\gL|^{-1}_{}\,\delta_{C_\gL}^\G\!(h_\la h^{-1}g)
  = |C_\gL|^{-1}_{}\,\delta_{C_{h_\la^{-1}\gL}}^\G\!(h^{-1}g) \,.
  \labl{DDD}


\paragraph{Simple current action.} 

Let us consider the action of \sicu s \J\ of the \gbar-\wzwm\ on the conjugacy 
classes $C^\G_\La\,{\subset}\,G$ that is obtained by mapping $C^\G_\La$ to
  \be
  C^\G_\La\cdo \gJ := \{ g \gJ \,|\, g \iN C^\G_\La \} \,\subset \G \,.
  \ee
Now the horizontal part of $\J\fus\Lax$ can be written as
(see e.g.\ section 14.2.2 of \cite{DIms})
  \be
  \J\fusb\La = w_{(\J)}w_\circ(\La) + \kg\LJ \,,
  \ee
where $w_\circ$ is the longest element of the Weyl group $W$ of \gbar\ and
$w_{(\J)}$ the longest element of the subgroup of $W$ that is generated by all
simple Weyl reflections except the one corresponding to the simple root that
is dual to \LJ. It follows that
  \be
  \J\fusb\La + \rho = w_{(\J)}w_\circ(\La{+}\rho) + (\kg\,{+}\,\gv)\, \LJ \,,
  \ee
and thus the group elements \erf{g-La} satisfy
  \be
  g^{}_{\J\fusb\La} = \big( w_{(\J)}w_\circ(g^{}_\La )\big)\,\gJ \,.
  \ee
Since Weyl transformations do not change the conjugacy class of a group 
element, this implies that
  \be
  C^\G_\La\cdo \gJ = C^\G_{\J\fusb\La} \,.
  \ee

Thus the \sicu\ group \Jg\ acts in a natural way via the fusion product on 
conjugacy classes of \G. Analogously, the \sicu\ group \Jh\ of the \hbar-\wzwm\
acts via the fusion product on conjugacy classes of \H, as
$C_\la^\H\,{\mapsto}\, C_\la^\H\cdo\hj \eq C_{\j\fusb\la}^\H$, and hence
\JgJh\ acts on the submanifolds \erf{CLala} of \GtH\ as
  \be 
  \CLala \,\mapsto\, C_{\J\fusb\La,\j\fusb\la} = \CLala\cdo (\gJ,\hj) \,.
  \ee
In particular, owing to $[g,h]\eq[gh^{-1},e]$ in \GHlr, it follows that
for $\gJ\eq\hj$ we have 
  \be
  \pilr(C_{\J\fusb\La,\j\fusb\la}) = \pilr(\CLala)
  \ee
for the coset branes \erf{pilrC}.
That is, if the \sicu\ $\Jj\iN\JgJh$ is an identification current,\,%
  \footnote{~This is the only generic reason for a \sicu\ $\Jj$ to act
  trivially on branes. On some specific branes, also \sicu s $\Jj\iN\JgJh
  {\setminus}\,\Id$ can act trivially, though.}
then it acts trivially on the coset branes \erf{pilrC}.


\section{Branes at finite level} \label{secBafl}

Like in any rational \cft, the symmetry preserving boundary states $\BQ{}$ 
of the coset theory are naturally labelled by the labels of primary fields, 
and when expressing them in terms of the Ishibashi functionals (or boundary
blocks) $\IQ{}$, which form bases of the spaces of conformal blocks for the
one-point correlators on the disk, the coefficients are given by the 
modular $S$-matrix. Thus, in the absence of field identification fixed points 
(which we assume), the $\BQ{}$ are labelled by the \Id-orbits $\lala$, and
are related to the Ishibashi states as
  \be
  \BQ\lala= \sum_{\lalaprime }\frac{S_{\lalao,\lalaoprime}}
  {\sqrt{S_{\lalaoprime,[0,0]}}}\, \IQ\lalaprime \,.
  \labl{BQ}
(Here and below we slightly abuse notation by writing $\La$ in place of
$\Lax$ in the subscripts of $S$-matrices.)
Also (again, in the absence of fixed points), the coset $S$-matrix is 
expressible through the modular $S$-matrices of the \gbar- and \hbar-\wzwm s
as
  \be
  S_{\lalaoprime,\lalao}
  = |\Id |\, S^\gbar_{\Lao,\Lao'}\, S^{\hbar~~*}_{\lao,\lao'} \,,
  \ee
where on the right hand side arbitrary representatives $(\Lax,\lax)$ and
$(\Lax',\lax')$ of the orbits $\lala$ and $\lalaprime$ are chosen 
\cite{gepn8,scya5}.

We would like to associate functions on the target space to the Ishibashi 
functionals $\IQ{}$, and thereby to the boundary functionals \erf{BQ}.
This allows one, via the approach of \cite{dfpslr,fffs}, to probe the geometry 
of a brane by bulk fields.\,%
  \footnote{~In the WZW case the analysis was done for bulk fields corresponding
  to, in string terminology, the graviton, dilaton and Kalb\hy Ramond field
  \cite{fffs}. To avoid having to deal explicitly with the coset chiral algebra,
  here we consider the bulk fields corresponding to the tachyon instead.
  As observed in \cite{fffs}, different choices of bulk
  fields give the same results qualitatively.}
For general background geometries it is a difficult task to find suitable
functions. But in the case of our interest a helpful strategy is to start
with a discussion of the branes on the group manifold $\GtH$, following the 
lines of \cite{fffs} for WZW branes. For group manifolds there is a preferred 
way of relating the functions on the target to the horizontal subspace 
$\Htorhor\,{\subset}\,\Htor$ of the space of bulk fields, and thereby to the 
boundary states which live in the dual space $\Htorhor^*$: the Peter\hy Weyl 
isomorphism \erf{P-W}, which associates to a vector 
$v\oti w \iN \big( \calh_\La^{}\oti\calh_\Lap{\big)}^*$ in the dual of 
\Htorhor\ the function 
  \be  
  f_{\rm PW}^{\gbar,\La}(v\oti w)
  \,{:=}\, \sqrt{{d_\La}/\VG}\, \langle v | D^\Lambda |w \rangle \,\in \FG\,.
  \ee
For any $\La$, the normalization of $f_{\rm PW}^{\gbar,\La}$ is determined, up 
to a phase, by the requirement that the mapping associates to vectors in an 
orthonormal basis of $\big( \calh_\La^{}\oti \calh_\Lap{\big)}^*$ functions 
that are orthonormal with respect to the Haar measure.

Using $\IG{\La'}(e^\La_m\oti e^\Lap_n)\eq\delta_{\La\La'}
\delta_{mn}$ (with $e^\La_m$ elements of an orthonormal basis of $\calh_\La^{}$,
as introduced after formula \erf{DccD} above), it follows that the function 
$\iG\La$ associated this way to the Ishibashi state $\IG\La$ is given by
  \be
  \iG\La(g)
  = \sqrt{{d_\La}/\VG}\,\sum_m \langle e^\Lap_m|\,D^\La(g)\,|e^\La_m \rangle
  = \sqrt{{d_\La}/\VG}\, \Chi_\La(g) \,,
  \labl{iG}
where $\Chi_\La$ is the character of the \G-\rep\ $R_\La$; we will refer
to this function as the {\em shape\/} of the Ishibashi state. Analogously, 
the Ishibashi states of the $\gbar{\oplus}\hbar$-\wzwm\ are given by
  \be
  \iGH\La\la(g,h) = \sqrt{\Frac{d_\La d_\la}{\VG\,\VH}}\,
  \Chi_\La(g)\, \Chi_\la(h)^*_{}
  = \sqrt{\Frac{d_\La d_\la}{\VG\,\VH}}\sum_{m,a} D^{\La,\la}_{mm,aa}(g,h)
  \,.
  \labl{iGH}
Here we have chosen a different convention on phases for the Peter\hy Weyl 
mapping $f_{\rm PW}^{\hbar,\la}$ of the \hbar-\wzwm, in order to conform with 
the fact that in the algebraic description instead of the \hbar-\wzwm\ it is 
the `complex conjugate theory' \hs\ that matters.

By comparison with the discussion of \FQ\ in section \ref{secGocm}, we expect
that in order to obtain the shape of Ishibashi states of the coset model, we 
should consider the projection 
  \be \bearl\dsty
  \sqrt{\Frac{d_\La d_\la\,\VH}{\VG}}\sum_{m,a} \pilrs D^{\La,\la}_{mm,aa}
  \\{}\\[-1.1em]\dsty\hsp4
  = \sqrt{\Frac {d_\La\,\VH} {d_\la^3\,\VG} }\,
  \sum_{\ij,\ik=1}^\multlala \sum_{m,p,q=1}^{d_\La} \sum_{a,b,c=1}^{d_\la}
  \cgcs\ij ma\la\La \cgc\ik ma\la\La\, \cgcs\ik qc\la\La \cgc\ij pb\la\La\,
  D^{\Lambda,\lambda}_{pq,bc}
  \\{}\\[-.8em]\dsty\hsp4
  =  \sqrt{\Frac {d_\La\,\VH} {d_\la\,\VG} } \,
  \sum_{\ij=1}^\multlala \sum_{p,q=1}^{d_\La} \sum_{b,c=1}^{d_\la}
  \cgcs\ij qc\la\La \cgc\ij pb\la\La\, D^{\Lambda,\lambda}_{pq,bc}
  \eear \labl{pDll}
of the functions \erf{iGH} to the coset manifold \Q\ (in the second equality, 
the unitarity of the matrices $c$ is used). However, a remark on the choice
of normalization made in \erf{pDll} is in order. In the WZW case, the prefactor 
in \erf{iG} is chosen in such a way that the integral $\intG\dg\bG{}(g)$ is of 
order $\kg^0_{}$ at large level.  Now whereas the normalization of the Ishibashi 
states $I^\G$ is unique up to a phase, the normalization of the functions $\iG{}$
\erf{iG} involves the one of the mappings $f_{\rm PW}^\gbar$.
The latter normalization is not intrinsic. In the WZW case we have fixed it
up to a phase by imposing orthonormality with respect to the Haar measure.
Similarly, for the \como\ we may require orthonormality of basis functions
in \FQ. Since our formulation starts with functions on $\G\,{\times}\,\HO$ 
rather than $\G/\HO$, this requires a compensating factor of $\VH$. 

Observe, however, that while the projections \erf{pDll} are well-defined 
functions on \Q, they are still labelled by (horizontally) allowed pairs 
$(\La,\la)$, whereas the coset Ishibashi states must be labelled by the allowed 
\Id-orbits $\lala$. Further, there is, in general, no bijection between allowed 
highest $\gbar{\oplus}\hbar$-weights and \Id-orbits of highest 
$\g{\oplus}\h$-weights. However, the selection rules are compatible in the 
sense that for 
every allowed pair $(\La,\la)$ the associated pair $(\Lax,\lax)$ belongs to an
allowed \Id-orbit. Moreover, inspection of various examples indicates that
every allowed \Id-orbit contains at least one representative whose horizontal
projection is a horizontally allowed pair, and that orbits with more than
one such representative are rare. (We are, however, not aware of any proof that 
this is actually true for all \como s.) Thus the projection still yields 
functions on \Q\ that closely match the properties of Ishibashi states, and 
accordingly we define the Ishibashi states as orbit sums of the functions
\erf{iGH}, i.e.\ set
  \be
  \iQ\lala([g,h]) := \sqrt{\Frac{\VH}{\VG}} \sum_{\Jj\in\Id} \!\!
  \sqrt{d_{\J\fusb\La} d_{\j\fusb\la}}\, \sum_{m,a}
  \pilrs D^{\J\fusb\La,\j\fusb\la}_{mm,aa}(g,h) \,.
  \labl{iQ}

The branes, i.e.\ the shapes of the boundary states, of the \como,
are then given by 
  \be
  \bearll  \bQ\lalaprime([g,h]) \!\! &= \dsty
  \sqrt{\Frac{\VH}{\VG}}
  \sum_\lala \frac{S_{\lalaoprime,\lalao}}{\sqrt{S_{\lalao,[0,0]}}}\, 
  \sum_{\Jj\in\Id} \!\! \sqrt{d_{\J\fusb\La}/d_{\j\fusb\la}}\,
  \\{}\\[-.8em] & \dsty \hsp{6.5} \sum_{\ij} \sum_{p,q} \sum_{b,c} 
  \cgcs\ij qc{\j\fusb\la}{\J\fusb\La} \cgc\ij pb{\j\fusb\la}{\J\fusb\La}\,
  D^{{\J\fusb\La},{\j\fusb\la}}_{pq,bc}(g,h) 
  \\{}\\[-.8em] & \dsty
  = {|\Id|}\, \sqrt{\Frac{\VH}{\VG}} \sum_\Lax
  \Frac {\sqrt{d_\La}\,S^\gbar_{\Lax',\Lax}} {\sqrt{S^\gbar_{\Lax,0}}} \,
  \sum_{p,q} D^\La_{pq}(g) 
  \sum_\lax \Frac {S^{\hbar~*}_{\lax',\lax}} {\sqrt{d_\la\,S^\hbar_{\lax,0}}}\, 
  \delta_{\la\prec\La} \sum_{\ij} D^{\la~~*}_{\till p\till q}(h) \,.
  \eear
  \labl{bQ}
Here in the second equality we switched to use adapted bases,
and combined the summation over allowed orbits and over the identification 
group to a summation over pairs $(\Lax,\lax)$. This yields of course only 
allowed pairs. But indeed we can sum over 
all pairs, including those which are forbidden by the selection rules,
because forbidden pairs are also horizontally forbidden and hence do not 
contribute owing to the presence of $\delta_{\la\prec\La}$.

\medskip

The formula \erf{bQ} is our result for the shape of symmetry preserving branes
in a \como\ without field identification fixed points. It expresses the coset 
branes entirely in terms of quantities for \g, \h\ and \G, \HO.
Notice that field identification is built in through the summation over the
identification group in \erf{iQ}. Still, in the formula \erf{bQ} the only 
explicit remnant of field identification is the overall
factor of $|\Id|$; such a simplification will certainly no longer arise in
models with field identification fixed points.

In the limit of large level the results
presented in this section reproduce those of the previous section.
Indeed one can also study the precise way in which the large level description 
emerges, but the analysis is somewhat involved and we refrain from going into
it in this paper. But anyhow it should be kept in mind that the limit of letting
the level approach infinity is not unique; some pertinent aspects of the large 
level limit will be discussed, in the context of WZW rather than \como s, 
in appendix \ref{appA}.

\medskip

There is an interesting class of \como s in which the formula \erf{bQ} 
simplifies considerably -- the case that $\HO\eq\T$ is a maximal torus of \G,
and hence $\hbar\eq\go$ the Cartan subalgebra of \gbar.
These \como s are known as {\em generalized parafermions\/} \cite{gepn2}.
It turns out that in this case the \rep\ matrices appearing in the formula
can be combined to characters of \G.
 
All irreducible \T-\rep s are \onedim, with \rep\ matrices the numbers $D^\la
(\eE^{\tcsa_\mu})\eq\eE^{(\la,\mu)}$. (Here we use the notation for Cartan 
subalgebra elements that was introduced before formula \erf{gJ}.)
The primary fields of the \hbar-theory
are labelled by the weight lattice of \gbar\ modulo \kg\ times the root lattice, 
so that their number $N_\T$ is $\kg^{\rm rank\,\gbar}$ times the number of \cc s 
of \gbar, i.e.\ $N_\T\eq\kg^{\rm rank\,\gbar}\,|\ZG|$. The modular S-matrix of 
the $\go$-theory is
  \be
   S^{\go}_{\lao',\lao} = N_\T^{-1/2} \eE^{-2\pi\ii\,(\la,\la')/\kg} \,,
  \ee
and the identification group \Ido\ is isomorphic as an abelian group to
the group \Jg\ of \sicu s of \gbar. It follows that
  \be
  \bearll  
  \bQ\lalaprime([g,h]) \!\!&\dsty 
  = |\Jg|\, N_\T^{-1/4}\, \sqrt{\Frac{\VT}{\VG}} \sum_\Lax
  \Frac {\sqrt{d_\La}\,S^\gbar_{\Lax',\Lax}} {\sqrt{S^\gbar_{\Lax,0}}} \,
  \sum_{\la\prec\La} \sum_\ij \eE^{2\pi\ii\,(\la,\la')/\kg} \,
  D^\La_{pp}(g)\, D^{\la\,*}(h)
  \\{}\\[-.8em] & \dsty
  = |\Jg|^{3/4}_{}\, \kg^{-\rm rank\,\gbar/4}\, \sqrt{\Frac{\VT}{\VG}}
  \sum_\Lax \Frac {\sqrt{d_\La}\,S^\gbar_{\Lax',\Lax}} {\sqrt{S^\gbar_{\Lax,0}}}
  \, \Chi_\La \big( \eE^{2\pi\ii\,t_{\la'}/\kg}\, h^{-1} g \big) 
  \,. \eear
  \labl{bGT}
Here in the first line $\ij$ labels the occurences of the weight 
$\la$ in $\La$, and we use the short-hand $p\,{\equiv}\,(\la;\ij)$. Thus
$\Chi_\La\eq\sum_p D^\La_{pp}$. The second equality holds because the
diagonal entries of the \rep\ matrices $D^\La$ satisfy
$D^\La_{pp}(gh)\eq D^\La_{pp}(g)\,D^\la(h)$ for any $g\iN\G$ and $h\iN\T$.


\section{Parafermions}

Parafermions can be realized by the $\sutwo_\kg / \uone_\kg$ 
coset construction \cite{DIms}.
The (horizontal part of) labels of the primary fields are pairs $(\La,\la)$ of 
integers in the range $0\,{\le}\,\La\,{\le}\,\kg$ and 
$-\kg\,{<}\,\la\,{\le}\,\kg$, subject to the field identification
  \be
  (\La,\la) \sim (\kg{-}\La,\la{+}\kg) 
  \ee
and the selection rule $2|\La{+}\la$, as well as $\la\,{\sim}\,\la{+}2\kg$
(stemming from the presence of an extended chiral algebra in the
$\uone_\kg$ theory). The identification group is a $\zet_2$ generated by the 
pair $(\kg,\kg)$. The nontrivial $\sutwo_\kg$ simple current $\La\eq\kg$ 
corresponds to the element $g_{(\kg)}\eq{-}\One \iN \G{\cap}\HO \eq \Uone$ 
with $\Uone \eq \{\eE^{\ii t \sigma_3} \,|\, t\iN [0,2\pi)\} \,{\subset}\,
\SUtwo$, while for the primary fields of $\uone_\kg$ (all of which are simple
currents) we have $h_{(\la)}\eq \eE^{\ii\pi\la\sigma_3/\kg}$.  

The representation matrices for \HO\ are the numbers
  \be 
  D^\la(\eE^{\ii t \sigma_3}) = \eE^{\ii \lambda t } \,.
  \ee
The horizontal branching spaces $\brsp_{\La,\la}$, where now
$\La\iN\zetpo$ and $\la\iN\zet$, are \onedim\ if $2|\La{+}\la$ and 
$|\la|\,{\le}\,\La$, and are zero else. Every orbit of the identification
group contains precisely one horizontally allowed pair, except for the
orbit of the identity field, for which both representatives $(0,0)$
and $(\kg,\kg)$ are horizontally allowed pairs.

\paragraph{Target space geometry.}
The target space of the parafermion theory can be described as follows. 
We parametrize, as in \cite{maMS}, the points on the manifold 
$\SUtwo\,{\cong}\, S^3$ as
  \be
  g = g(\psi,\theta,\phi) = \exp(\ii\psi \sigma_{\vec n})
  = \cos\psi\,\One + \ii\,\sin\psi\, \sigma_{\vec n} \,,
  \labl{gpsi}
where $\vec n$ is a point on the unit two-sphere with the standard
coordinates $\theta$ and $\phi$, and $\sigma_{\vec n}\,{:=}\,
\vec n\,{\cdot}\,\vec\sigma$ with $\vec\sigma\eq(\sigma_1,\sigma_2,\sigma_3)$
the Pauli matrices. Then the coordinate ranges are 
$\psi\iN [0,\pi]$, $\theta\iN [0,\pi]$ and $\phi\iN [0,2\pi)$, and taking the
radius of the three-sphere to be $\sqrt\kg$, the metric is
 $ {\rm d}s^2 \eq \kg\,({\rm d}\psi^2\pl\sin^2\!\psi$\linebreak[0]$
 {\rm d}\theta^2 \pl\sin^2\!\psi\,\sin^2\!\theta\,{\rm d}\phi^2) $,  
so that $\VG\eq2\pi^2\kg^{3/2}$ and $\VH\eq2\pi\kg^{1/2}$.
The non-trivial simple current of the \sutwo\ theory acts on \SUtwo\ as 
  \be
  g = g(\psi,\theta,\phi) \;\mapsto\;
  g\,\gJ = -g = g(\pi{-}\psi,\pi{-}\theta,\pi{+}\phi) \,,
  \ee

\begin{figure}[th] 
\begin{center}
\begin{picture}(170,110)(0,9)
    \put(21,9)     {\begin{turn}{-5}\scalebox{.4}
                   {\includegraphics{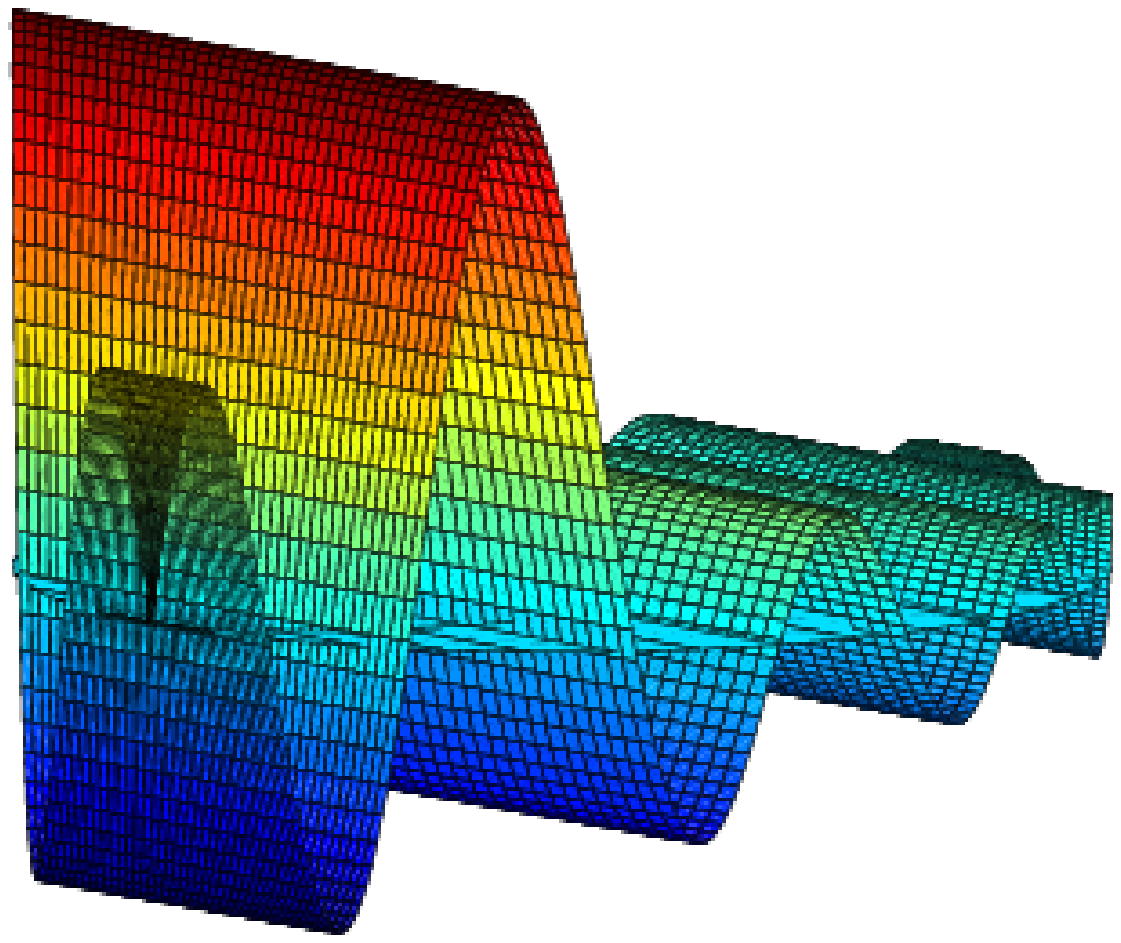}}\end{turn}}
    \put(-28,-9)   {\scalebox{.4}{\includegraphics{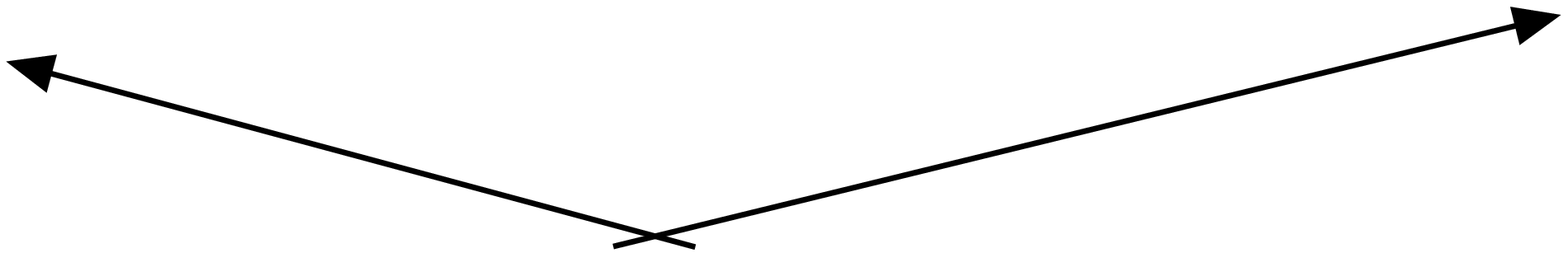}}}
    \put(-36.6,16.2) {\small\begin{turn}{78}$y$\end{turn}}
    \put(192.6,27.3) {\small\begin{turn}{296}$x$\end{turn}}
\end{picture}\end{center}
\caption{The shape $\bPF\lala$ of the parafermion brane
         for $(\La,\la)\eq(8,0)$ at level $\kg\eq10$.}  
\label{fig:disc}\end{figure} 

~\vfill
\noindent
and the simple currents $\la$ of the \uone\ theory act on the maximal torus
as multiplication by $\eE^{\ii\pi\la\sigma_3/\kg}$, i.e.\ as a rigid rotation
by the angle $\la\pi/\kg$.

The subgroup $\HO\eq\Uone$ whose adjoint action is gauged is the maximal torus,
which in the parametrization \erf{gpsi} is given by
  \be
  \HO = \{ \eE^{\ii t \sigma_3} \,|\, t\iN[0,2\pi) \}
  = \{ g(\psi,\theta,\phi) \,|\, \phi\eq0,\, \theta\eq0,\pi \} \,.
  \ee
Conjugation by $\eE^{\ii t \sigma_3}$ amounts to a shift in the 
$\phi$-coordinate, and hence the parafermion target space \QPF\
is parametrized by $\psi\iN [0,\pi]$ and $\theta\iN [0,\pi]$. 
In terms of the coordinates 
  \be
  x := \Frac\pi2\, \cos \psi  \qquad{\rm and}\qquad
  y := -\Frac\pi2\, \cos\theta\, \sin \psi \,,
  \labl{xy}
\QPF\ is the set of points $(x,y)\iN\reals^2$ in the range 
$x\iN\big[-\Frac\pi2,\Frac\pi2\big]$, $y\iN\big[
{-}\sqrt{(\frac\pi2)^2\,{-}\,x^2}\,, \sqrt{(\frac\pi2)^2\,{-}\,x^2}\,\big]$,
i.e.\ a (round) disk of diameter $\pi/2$. Embedding this disk \Dpi\ in its 
covering space \SUtwo\ at $\phi\eq0$ describes it as the set of group elements
  \be 
  g\,(x,y) \equiv g \big(\! \arccos\Frac{2x}{\pi},
  \arccos\Frac{-2y}{\sqrt{\pi^2-4x^2}} , 0 \big)
  = \Frac{2x}\pi\,\One - \Frac{2y}\pi\,\ii\sigma_3 
  + \Frac1\pi \sqrt{\pi^2{-}4x^2{-}4y^2}\,\ii\sigma_1 \,. 
  \labl{gxy} 

\paragraph{Brane geometry.}
The parafermions constitute the simplest case $\G\eq\SUtwo$ of the
$\HO\eq\T$ situation, in which the formula \erf{bQ} reduces to \erf{bGT}.
Plugging in the data for the parafermions into \erf{bGT}, we obtain 
  \be
  \!\bearll  \bPF\lalaprime([g,\expits]) \!\! &= \dsty
  \Frac {|\Id|\,\sqrt{\VH}} {(\kg(\kg{+}2))^{1/4}\, \sqrt{\VG}}
  \sum_{\La=0}^{\kg}
  \Frac { \sqrt{\La+1}\, \sin\big((\La+1)(\La'+1)\Frac\pi{\kg+2}\big) }
        { \sqrt{ \sin\big((\La+1)\Frac\pi{\kg+2}\big) } } 
  \sum_{\stackrel{\scriptstyle \la=-\La}{2|\La+\la}}^{\La}\!
  \eE^{\ii\pi \la\la'/\kg} D^\La_{\la\la}(g) \, \eE^{-\ii\la t} 
  \\{}\\[-.8em] &\dsty
  = \Frac {2} {\sqrt{\pi\,\kg}\,(\kg(\kg{+}2))^{1/4} }\, \sum_{\La=0}^{\kg}
  \Frac { \sqrt{\La+1}\, \sin\big((\La+1)(\La'+1)\Frac\pi{\kg+2}\big) }
        { \sqrt{ \sin\big((\La+1)\Frac\pi{\kg+2}\big) } } \,
  \Chi_\La \big( \eE^{ \ii\sigma_3 (-t + \pi\la'/\kg)} g \big) \,.  
  \\[-4.0em]~ \eear
  \labl{bPF}

\begin{figure}[tb] 
\begin{center}\begin{picture}(400,150)
    \put(-33.3,12) {\scalebox{.22}{\includegraphics{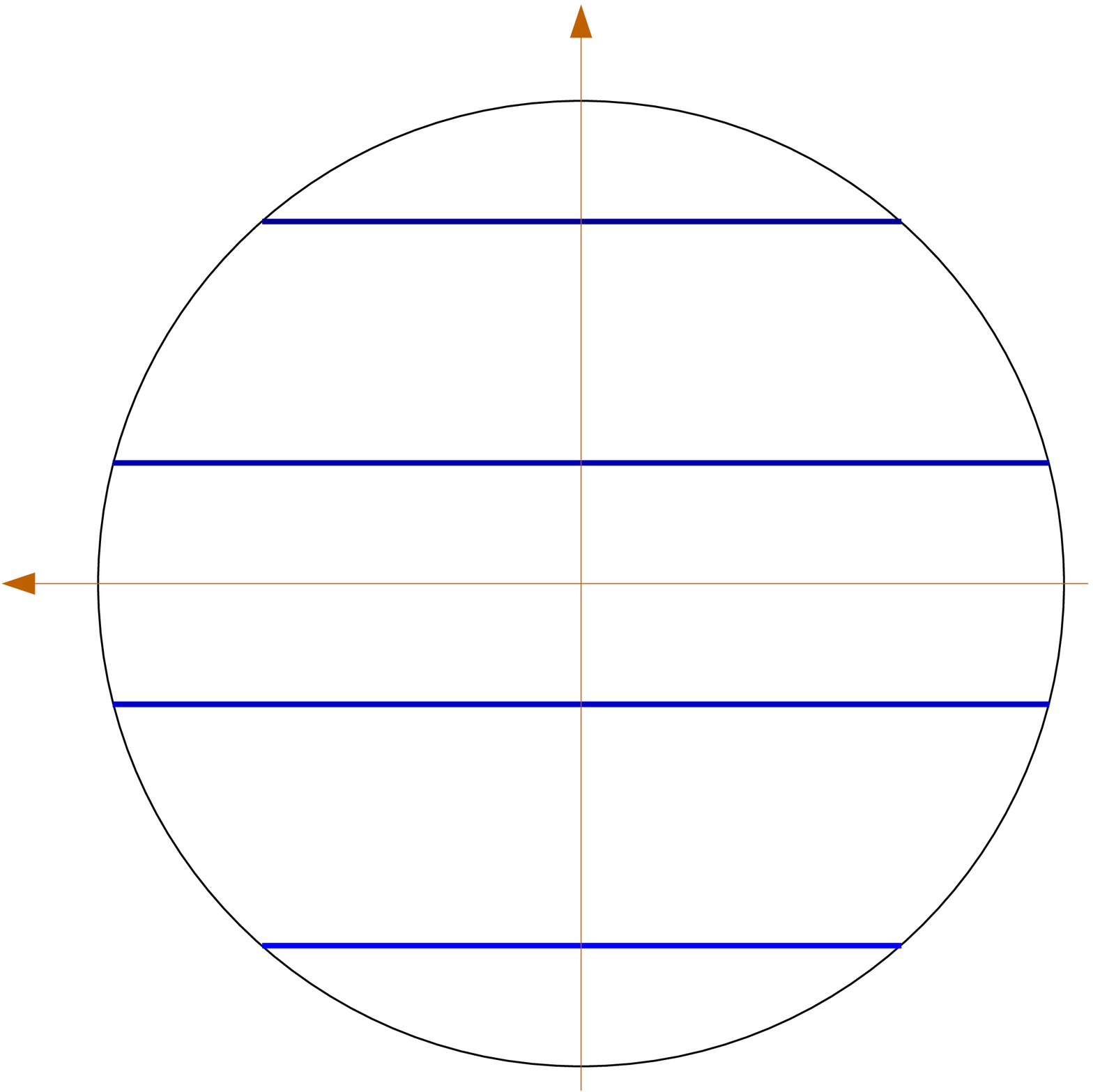}}}
    \put(-25,12) {\begin{picture}(0,0)
                 \put(51,-20) {(a)}
                 \put(56.7,128.7) {\scriptsize$x$}
                 \put(-15.2,57.2) {\scriptsize$y$}
                 \put(100,99) {\scriptsize(0,0)}
                 \put(114,71) {\scriptsize(2,0)}
                 \put(114,43) {\scriptsize(4,0)}
                 \put(100,15) {\scriptsize(6,0)}
                 \end{picture}}
    \put(136.7,12) {\scalebox{.22}{\includegraphics{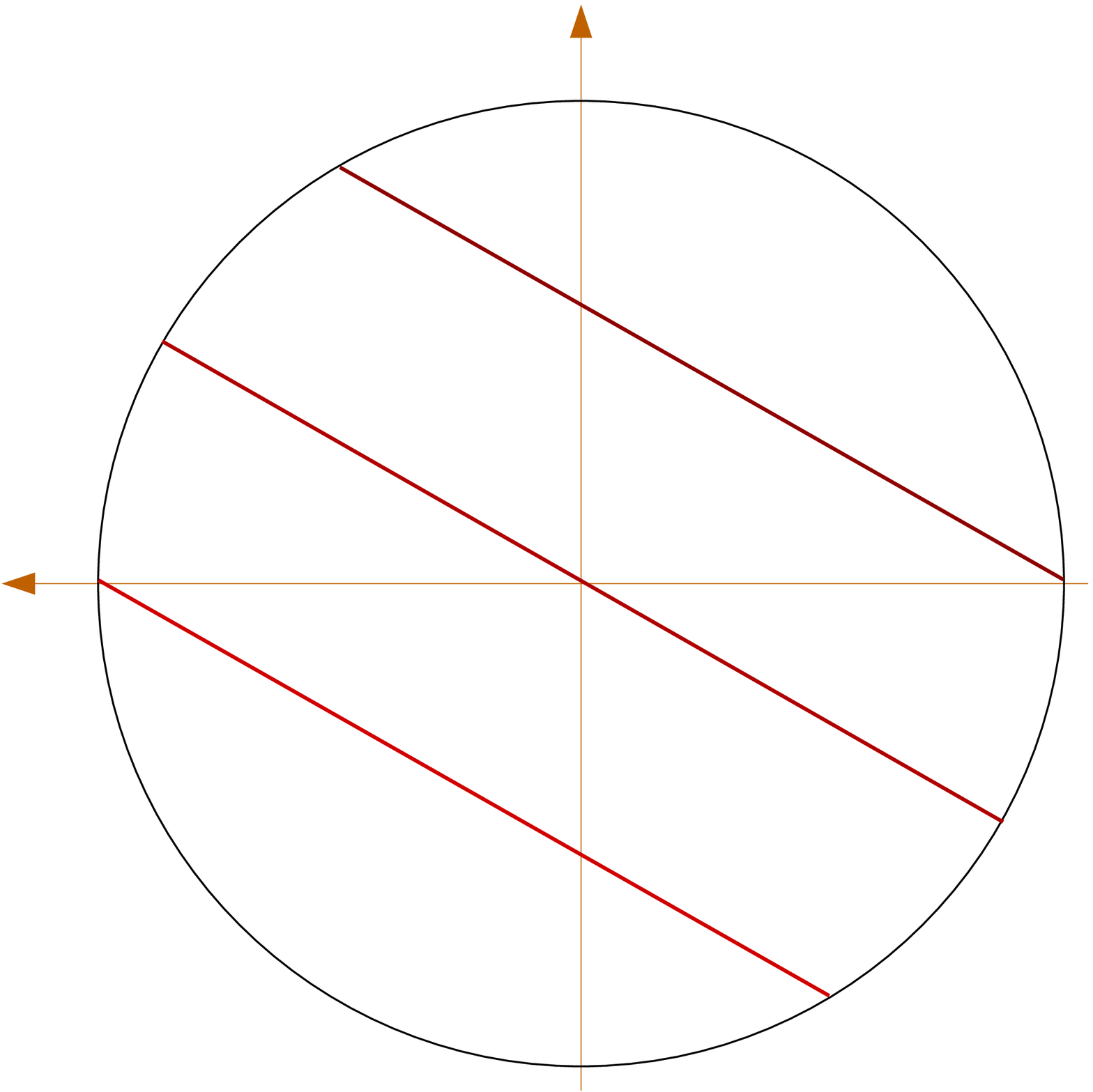}}}
    \put(145,12) {\begin{picture}(0,0)
                 \put(51,-20) {(b)}
                 \put(56.7,128.7) {\scriptsize$x$}
                 \put(-15.2,57.2) {\scriptsize$y$}
                 \put(12,108) {\scriptsize(1,1)}
                 \put(109,29) {\scriptsize(3,1)}
                 \put(87,4)   {\scriptsize(5,1)}
                 \end{picture}}
    \put(306.7,12) {\scalebox{.22}{\includegraphics{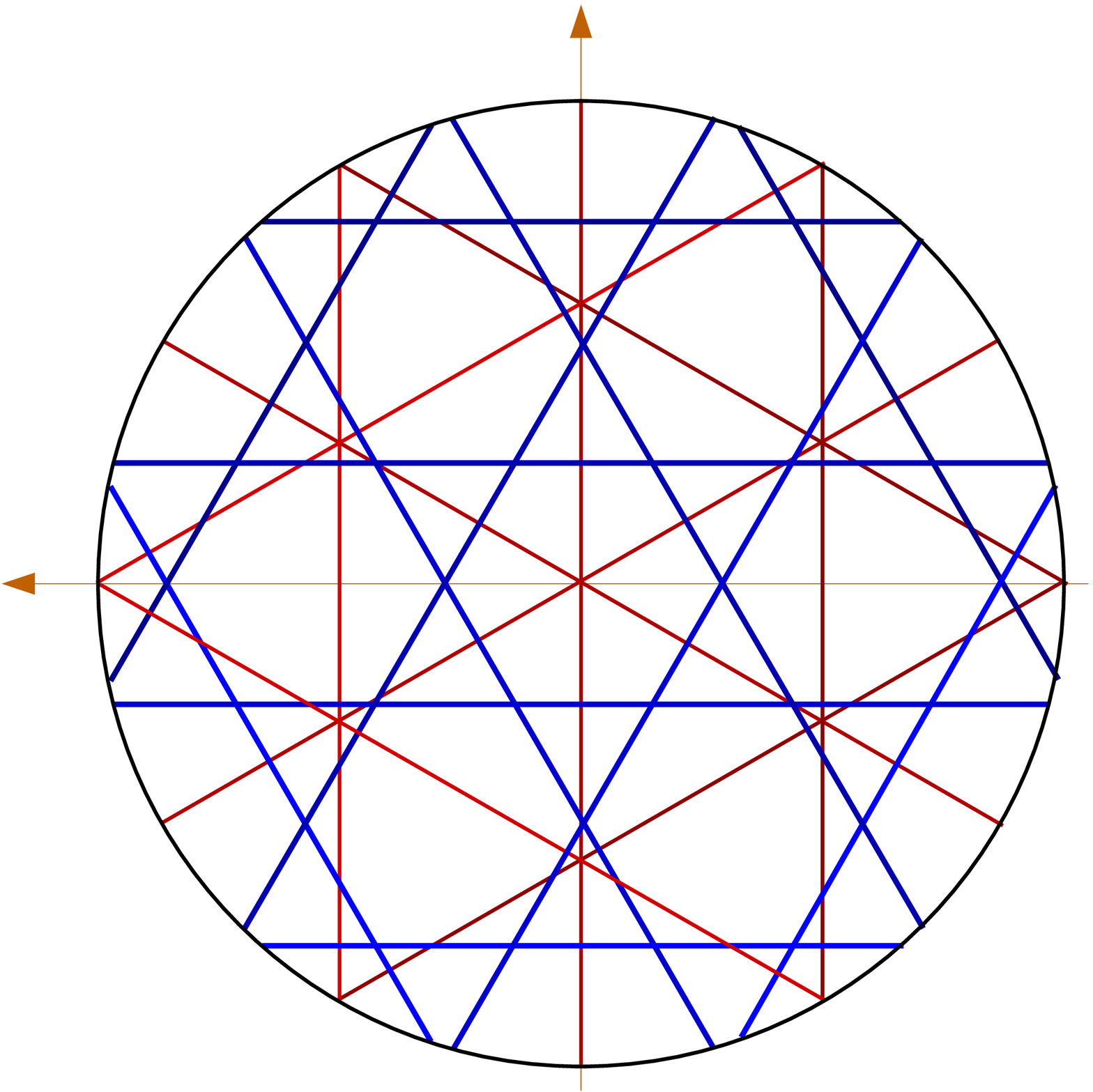}}}
    \put(315,12) {\begin{picture}(0,0)
                 \put(51,-20) {(c)}
                 \put(56.7,128.7) {\scriptsize$x$}
                 \put(-15.2,57.2) {\scriptsize$y$}
                 \end{picture}}
\end{picture}\end{center}
\caption{The straight lines on $\QPF\eq\Dpi$ at which the parafermion branes at
  level $\kg\eq6$  are centered:%
  ~~(a)~The branes $(\La,0)$.~~(b)~The branes $(\La,1)$.~~(c)~All 21 branes.~~
  Note that some of the branes meet on the boundary $\partial\Dpi$, but that 
  this is not the generic situation.}  
\label{fig:linesAll}\end{figure} 

\begin{figure}[bh] 
\begin{center}
\begin{picture}(300,265)
    \put(-43,-14)    {\scalebox{.48}{\includegraphics{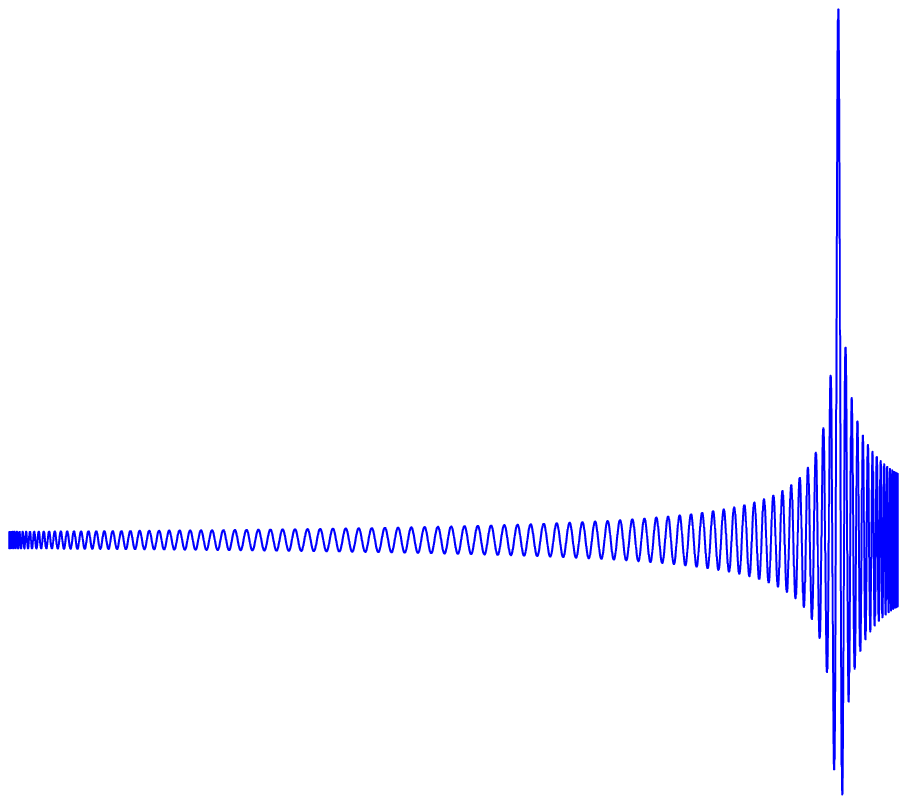}}}
    \put(136,-20)    {\scalebox{.48}{\includegraphics{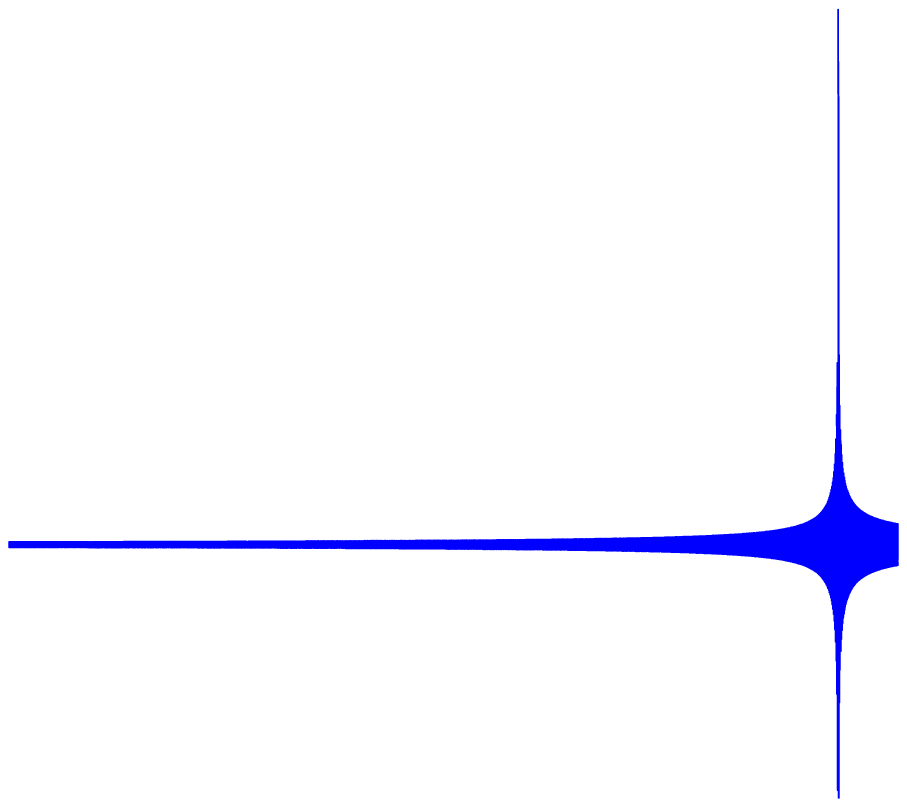}}}
    \put(-43,140.3)  {\scalebox{.48}{\includegraphics{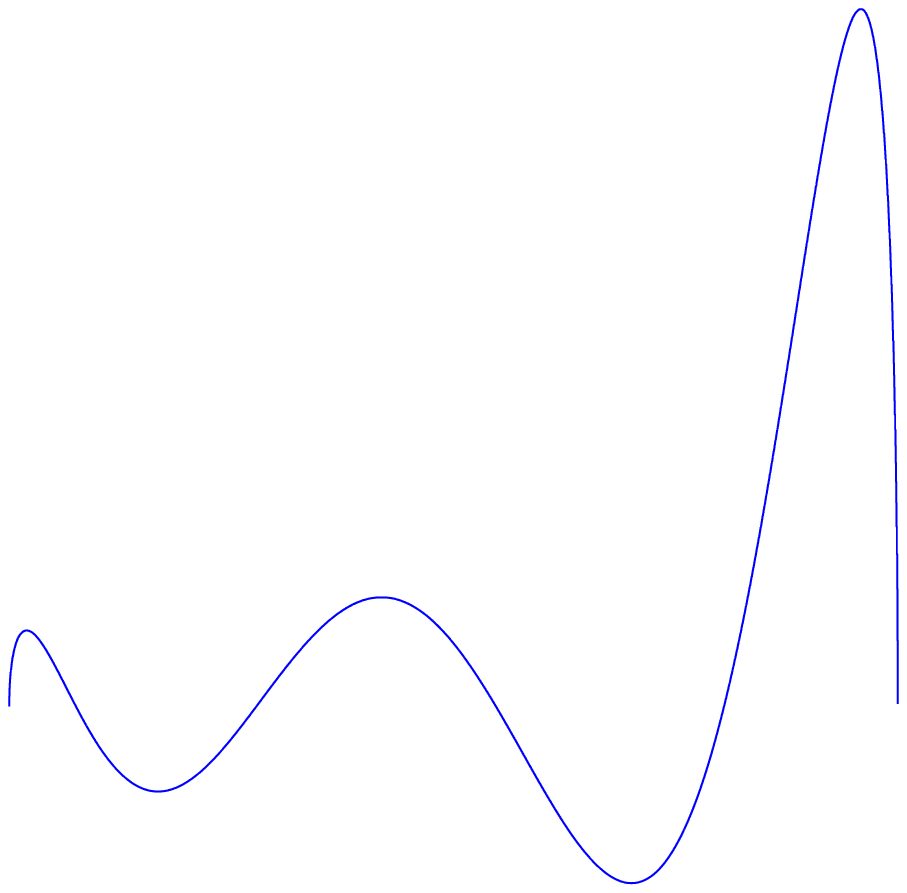}}}
    \put(136,132.5)  {\scalebox{.48}{\includegraphics{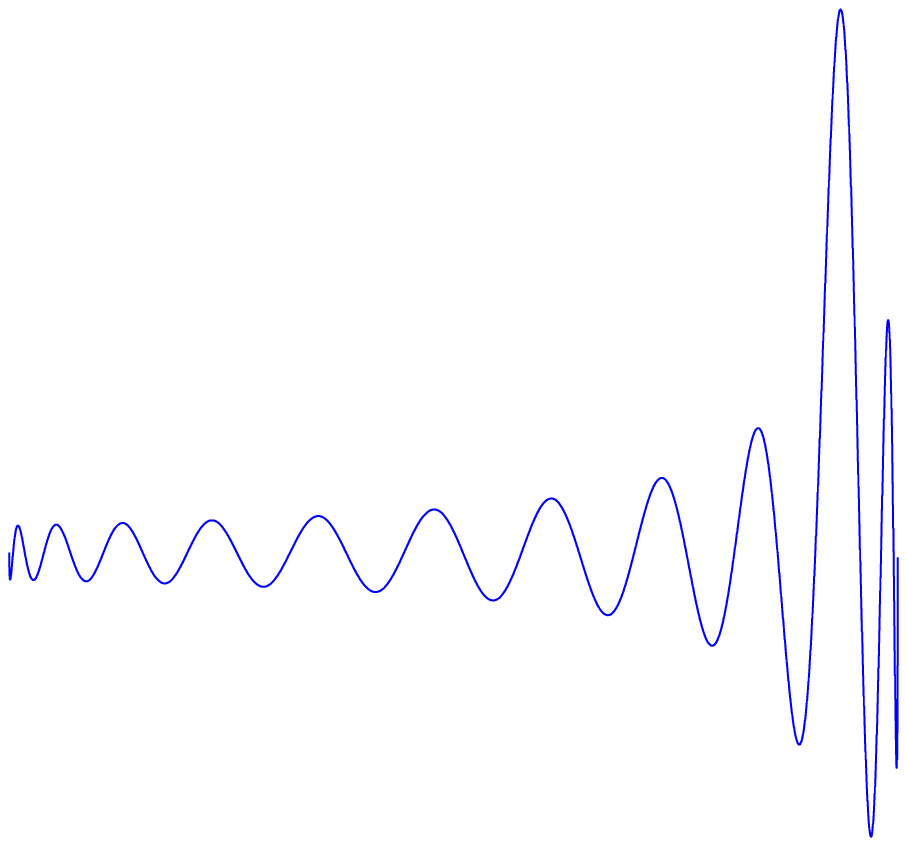}}}
    \put(-2,150) {\scalebox{.246}{\includegraphics{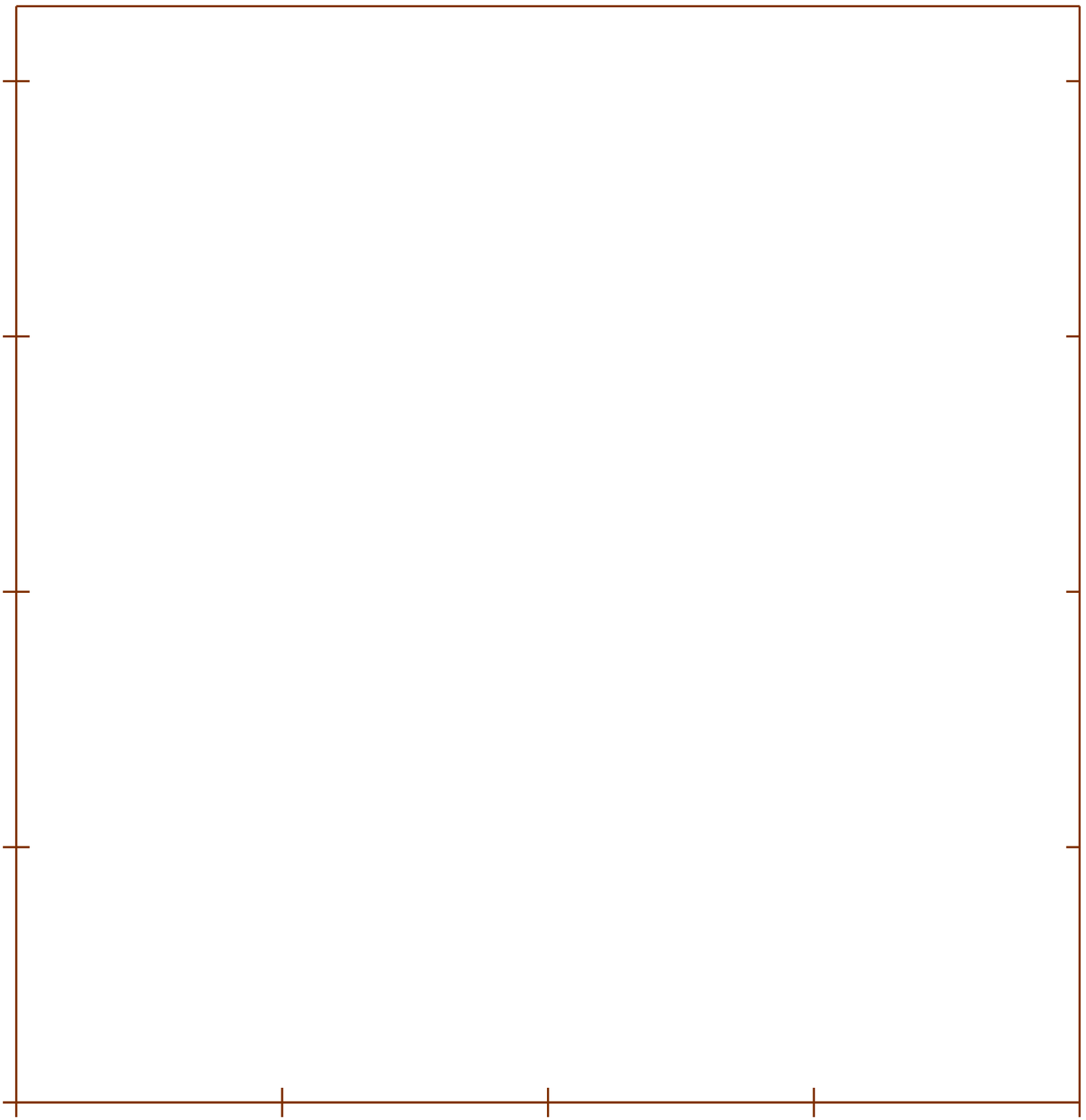}}}
    \put(0,150)  {\xyaxisulevel 4 \begin{picture}(0,0)(3.2,0)
                   \put(-15.4,.5)  {\tiny$-.5$}
                   \put(-7.1,30.2) {\tiny$0$}
                   \put(-9.9,60.2) {\tiny$.5$}
                   \put(-7.2,89.8) {\tiny$1$}
                   \put(-14.3,119.5){\tiny$1.5$} \end{picture} } 
    \put(176,150){\scalebox{.246}{\includegraphics{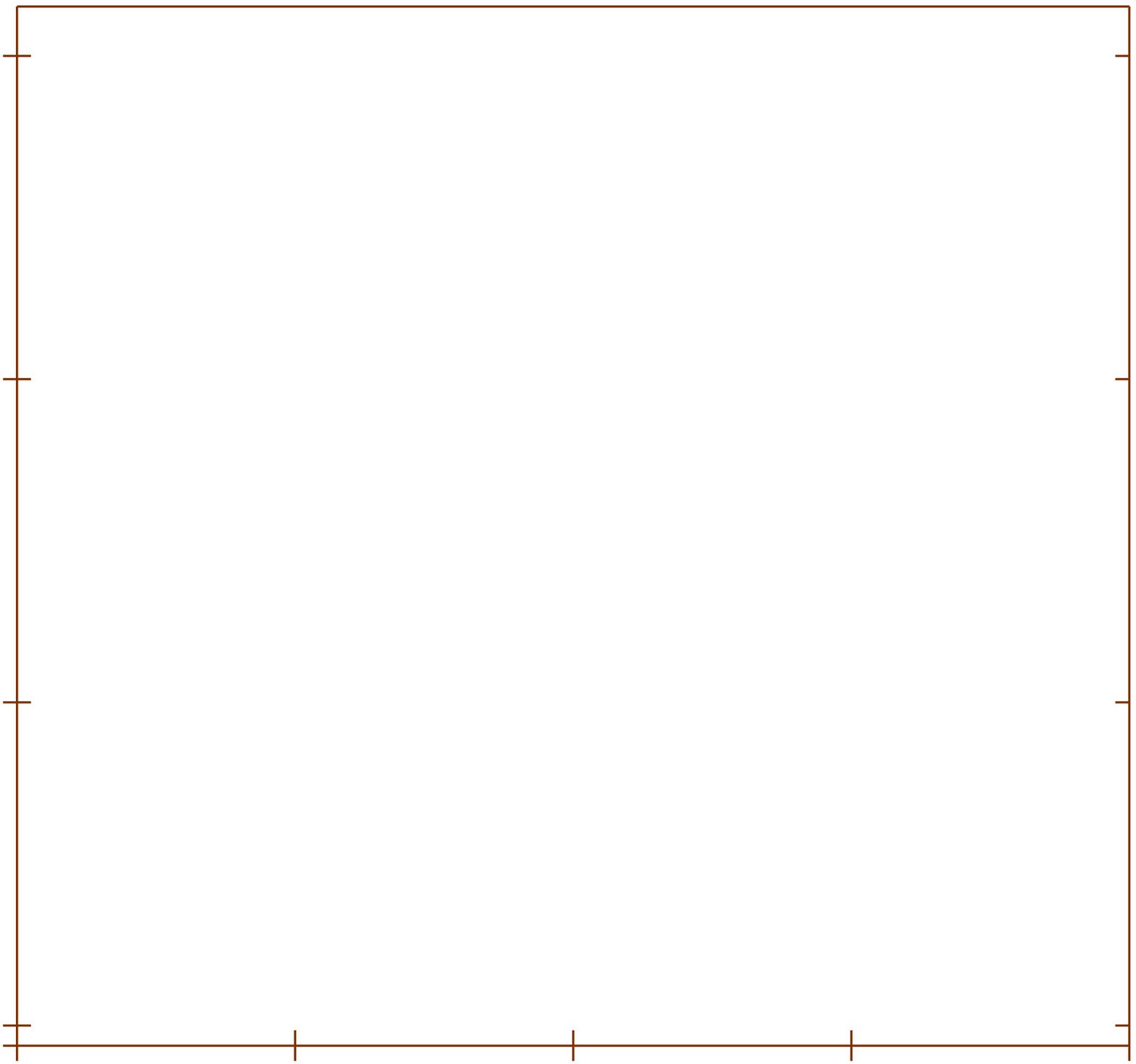}}}
    \put(178,150){\xyaxisulevel {22} \begin{picture}(0,0)(3.2,0)
                   \put(-12.4,2.6) {\tiny$-1$}
                   \put(-6.9,38.5) {\tiny$0$}
                   \put(-7.1,74.5) {\tiny$1$}
                   \put(-7.2,111)  {\tiny$2$} \end{picture} }
    \put(-2,-2)  {\scalebox{.246}{\includegraphics{PFcxes.eps}}}
    \put(0,-2)   {\xyaxisulevel {208} \begin{picture}(0,0)(3.5,0)
                   \put(-12.4,2.6) {\tiny$-4$}
                   \put(-6.6,38.2) {\tiny$0$}
                   \put(-7.1,74.5) {\tiny$4$}
                   \put(-7.1,111)  {\tiny$8$} \end{picture} }
    \put(176,-2) {\scalebox{.246}{\includegraphics{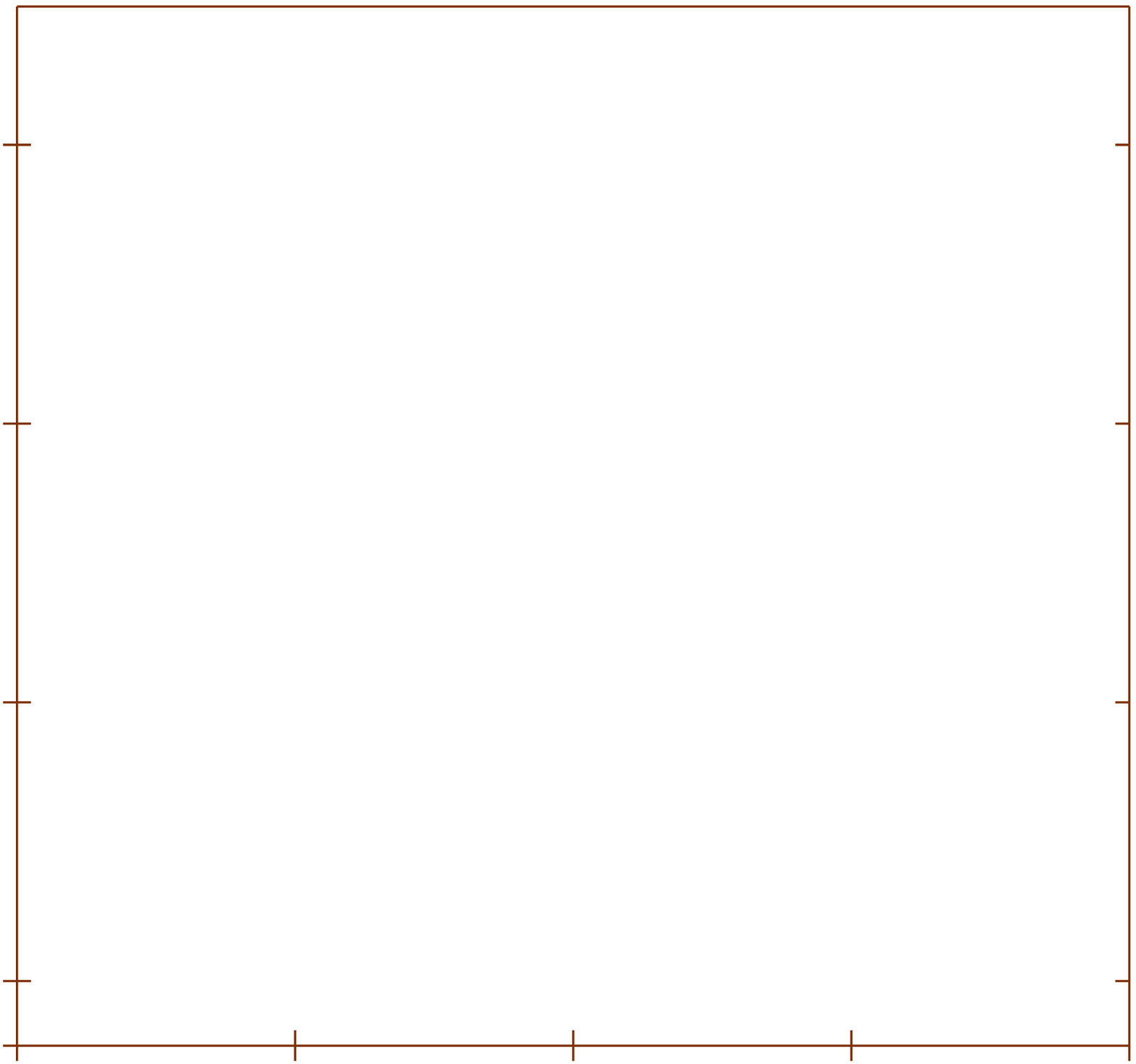}}}
    \put(178,-2) {\xyaxisulevel {2098} \begin{picture}(0,0)(3.2,0)
                   \put(-17,7.9)   {\tiny$-10$}
                   \put(-6.6,38.2) {\tiny$0$}
                   \put(-11.5,69.7){\tiny$10$}
                   \put(-11.5,101) {\tiny$20$} \end{picture} }
\end{picture}
\end{center}
\caption{The shape $\tbPF{}$ of the brane $(\frac{\kg-4}6,0)$ as a 
function of the coordinate $x$, at four different values of the level.
At large level, the brane is concentrated at $x\eq\frac{5\pi}{12}$.}
\label{fig:four}
\end{figure}

\vfill
\clearpage

\noindent
While the \rhs\ of \erf{bPF} is written as a function of $g'\eq\eE^{\ii\sigma_3
(-t+\pi\la'/\kg)}g$, i.e.\ like a function on the group \SUtwo,
it is indeed a function on \QPF, since $g'(x,y)$ as given by \erf{gxy} just
projects to $(x,y)\iN\Q$. Also, as the group characters $\Chi_\La$ depend only 
on a single variable, each brane actually depends on a definite combination of 
$x$ and $y$. Concretely, with our choice of coordinates they are constant 
along some direction on the disk. This is most directly seen for branes with 
$\la\eq0$; they depend only on $x$, but not on $y$. Further, 
the factor $\eE^{\ii\pi\lambda\sigma_3/\kg}\iN\Uone$ amounts to a rigid rotation
of $\partial\Dpi$ by an angle $\pi\lambda/\kg$ about its center, so that the
straight line for the brane labelled by $(\La,\la{+}\la')$ is obtained from 
the one for the brane $(\La,\la')$ by such a rotation (see also \cite{fred'}).
We illustrate this behavior in figure \ref{fig:disc} for a brane at level 10.
The same observation also shows that the branes are mapped to themselves by
the action of the non-trivial identification current.

In the large level limit, according to \erf{DDD} the shape of each brane 
converges to a multiple of the delta function on the projection of the product 
of the relevant conjugacy classes, i.e.\ of the class of $\expits g$ in \SUtwo\ 
and of the point $\eE^{\ii\pi\lambda\sigma_3/\kg}\iN\Uone$. By the previous 
remarks, this yields just a delta function on a straight line in \Dpi;
these lines are shown, for level $\kg\eq6$, in figure \ref{fig:linesAll}.
 
At finite level the branes are peaked along these straight lines, but they 
are smeared significantly about these subsets. As an illustration, in figure 
\ref{fig:four} we display the shape of the brane $(\La,0)$ as it evolves with 
the level, plotted as a as a function of the coordinate $x$, for 
$\La\eq\La(\kg)$ chosen such that $\La(\kg){+}1\eq(\kg{+}2)/6$.
Since we draw the shapes as a function of the variable along the straight
line, rather than on the disk, we must account for the different extension in
perpendicular direction by an approapriate measure factor; in the case at hand,
this amounts to replacing $\bPF{}(x)$ by $\tbPF{}(x)\eq
2(\frac{\pi^2}4\,{-}\,x^2)^{1/2}_{}\,\bPF{}(x)$.


\appendix

\section{The large level limit}\label{appA}

In the limit of large level the world volumes of the (symmetry preserving)
coset branes -- the subsets of \Q\ on which the branes are concentrated -- 
are lower-dimensional submanifolds of \Q. When performing the limit for 
our result \erf{bQ}, one is led to conclude that these submanifolds 
have the same dimension for all branes. For a more explicit description, 
the particular way of taking the large level limit must be specified, however.  
Rather than performing this analysis for \como s, in this appendix we 
present the corresponding results for \wzwm s in some detail. The coset case
can then be treated analogously, but requires more notational complexity.
As already mentioned, there is no unique way of taking the limit $\kg\,{\to}\,
\infty$. That one better does not draw conclusions about the large level 
behavior too quickly is already apparent from the observation that naively 
at $\kg\,{\to}\,\infty$ the target space is a group manifold with infinite 
radius, which superficially looks as flat space even though it should
still be compact.

One possibility is to consider the situation that one deals with a definite
conjugacy class $C$. This requires to let the weight $\La$ labelling a brane 
depend on the level in such a manner that the associated group elements $g_\La$ 
\erf{g-La} belong to the desired conjugacy class $C$. This limit has been 
already described in detail in \cite{fffs}. (The analogous procedure for 
parafermions amounts to a level dependence of the type occuring in figure 
\ref{fig:four}.) In the WZW case one finds that, analogously as at any finite 
level, in the limit all brane world volumes are concentrated on {\em
regular\/} conjugacy classes, and already at small level the overlap
with the exceptional lower-dimensional conjugacy classes is negligible. 

\smallskip

An alternative limit consists in keeping instead the weight $\La$ fixed.
It is easily seen that in this case at large levels for {\em any\/} $\La$ the 
position of the center of the brane is driven to the unit element of \G, 
i.e.\ to an execptional conjugacy class.  In other words, each brane tends 
to a D0-brane. On the other hand, upon closer inspection it turns out
that even in this limit all branes are, in a specific sense, still well
separated from the unit element. The rest of this appendix is devoted to 
show how one arrives at this conclusion.

\begin{figure}[tb] 
\begin{center}
\begin{picture}(230,110)
    \put(0,0)     {\scalebox{.6}{\includegraphics{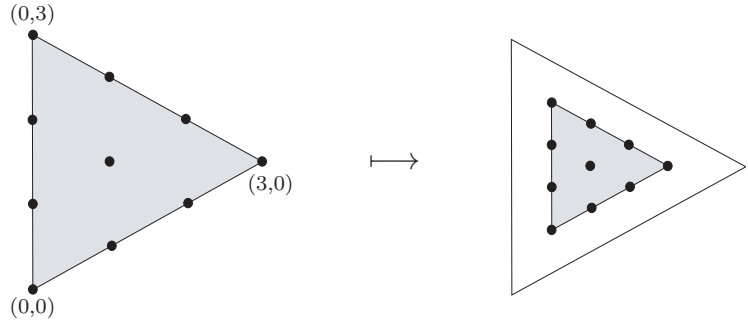}}}
    \put(-7,-6.6) {\scriptsize(0,0)}
    \put(-7,104.4){\scriptsize(0,3)}
    \put(83,40)   {\scriptsize(3,0)}
    \put(129,48)  {$\longmapsto$}
\end{picture}
\end{center}
\caption{The mapping \erf{shift} of the fundamental Weyl alcove $\mathscr W$ 
for \suthree\ at level 3. The left picture shows the alcove $\mathscr W$,
with  the dots indicating the location of the integral weights, while
the right picture shows its image $\tilde{\mathscr W}$ under \erf{shift} inside 
the original region. The points on the boundary $\partial\tilde{\mathscr W}$
are those weights $\tilde\La$ whose associated Cartan subalgebra elements are
mapped by the exponential mapping to points on exceptional conjugacy classes of 
\SUthree.} 
\label{fig:weyl4}
\end{figure}

First notice that the result that at finite level all branes are concentrated 
at regular conjugacy classes has its origin in the fact that
it is not the integrable highest weights $\La$ that specify the locations of 
the conjugacy classes on the maximal torus of \G, but rather there is a 
shift combined with a compression,
  \be
  \Lambda \,\mapsto\, \tilde \Lambda \,:= \Frac{\kg}{\kg+\gv}\,(\La+\rho) \,,
  \labl{shift}
which maps the fundamental Weyl alcove to its interior.
Since with increasing \kg\ the shift $\kg\,{\mapsto}\,\kg{+}\gv$ is more and
more difficult to detect, certain semiclassical formulas ignore the modification
\erf{shift}, and hence this aspect of the branes is missed when applying such
formulas directly to the finite-level situation.
The absence of exceptional conjugacy classes is particularly significant
at low levels, at which the fraction of dominant integral weights lying on 
the boundary of the fundamental alcove is large; this is illustrated with
an \SUthree\ example in figure \ref{fig:weyl4}.

\smallskip

Let us now present for fixed $\La$ the large-level behavior of the boundary 
coefficients $\BG\La$ that appear in the expansion of boundary states in 
terms of Ishibashi states, i.e.\ in the WZW analogue of the formula \erf{BQ}.
We use the the fact that $S_{\Lao',\Lao}/S_{\Lao,0}\eq\Chi_{\La'}(g_\La^{-1})$
(see e.g.\ formula (13.8.9) of \cite{KAc3}), where $g_\La$ is 
the group element associated to the weight $\La$, see \erf{g-La}, and that 
according to 
  \be
  \frac{S_{0,\Lao}}{S_{0,0}} = \frac
  {\prod_{\alpha\in\Phi}\sin\big( \frac{(\La+\rho,\alpha)}{\kg+\gv}\, \pi \big)}
  {\prod_{\beta\in\Phi}\sin \big( \frac{(\rho,\beta)}{\kg+\gv}\, \pi \big)}
  ~~\stackrel {\raisebox{2pt}{$\scriptstyle\kg\to\infty$}}\longrightarrow~~
  \prod_{\alpha\in\Phi} \frac{(\Lambda{+}\rho,\alpha)}{(\rho,\alpha)} = d_\La 
  \ee
with $\Phi$ the set of positive roots of \gbar, at large level the quantum 
dimensions approach the ordinary dimensions.  
Following the arguments in \cite{fffs}\,%
  \footnote{~Apart form the different normalization convention for Ishibashi
  states, a factor of $S_{0,\Lao}/S_{0,\Lao'}$ is lacking on the \rhs\ of
  formula (2.4) of \cite{fffs}. This does not effect the qualitative behavior
  of the branes.}
we then obtain
  \be
  \frac {S_{\Lao',\Lao}}{\sqrt{S_{0,\Lao'}}}
  \,\longrightarrow\, \sqrt{S_{0,0}} \,\frac{d_\La}{\sqrt{d_{\La'}}}\,
  \Chi_{\La'} (g_\La^{-1}) \,.
  \ee
It is here that the shift \erf{shift} enters the story.

Implementing the orthogonality and completeness of the characters, it follows 
that in the large level limit the shape of the branes behaves as
  \be
  \bG\La(g) \,\longrightarrow\, d_\La\, \sqrt{\Frac{S_{0,0}}{\VG}} \sum_{\La'}
  \Chi_{\La'}(g_\La^{-1})\,\Chi_{\La'}(g)
  = d_\La\, \sqrt{S_{0,0}\,\VG}\,\frac{\delta_{C_\gL}^\G\!(g)}{|C_\gL|}\,,
  \labl{WZWB}
where $C_\gL{\subset}\,\G$ is the conjugacy 
class of $g_\La$ and $\delta_C$ is the class delta distribution, which
acts on class functions $f$ as $\int_{\!\G}{\rm d}u\,\delta_{C_g}^\G(u)\,f(u)
\eq |C_g|\,f(g)$ for $g\iN\G$.
Note that $\delta_{C_\gL}^\G/|C_\gL|\eq\delta_\gL^\T$ is nothing but a 
delta distribution on the maximal torus of \G, and that in the limit 
$\kg\,{\to}\,\infty$ the product $S_{0,0}\,\VG$ approaches a \kg-independent 
constant. Thus in short, for fixed $\La$ in the large level limit the shape 
of the brane labeled by $\La$ is a constant multiple of a delta function 
concentrated at $g_\La$ on the maximal torus.  

For continuing the discussion, let us restrict our attention to $\G\eq\SUtwo$.

\smallskip

In the $\sutwo_\kg$ WZW model, the horizontal weights labelling the branes are 
the integers in the range $0\,{\le}\,\La\,{\le}\,\kg$. 
Parametrizing $g\iN\SUtwo$ as in \erf{gpsi}, the branes are given by
  \be
  \bZ\Lax(g) 
  = \frac{1}{\sqrt {|\SUtwo|}}\;{^{4}\!\!\!\sqrt{\Frac{2}{\kg+2}}}
  \,\sin^{-1}\!\psi\sum_{\La'=0}^\kg \!\sqrt{\La'{+}1}\,
  \frac{\sin\big((\La'{+}1){\psi}_\La\big)\,\sin\big((\La'{+}1)\psi\big)}
  {\sqrt{\sin\big(\frac{(\La'+1)\pi}{\kg+2}\big)}}
  \labl{bZ2}
with
  \be {\psi}_\La := \frac{(\La\,{+}\,1)\,\pi}{\kg\,{+}\,2} \,, 
  \labl{psiLa}
compare formula (D.4) of \cite{maMS}. 
According to \erf{WZWB} this function behaves for large \kg\ as
  \be
  \bZ\Lax \longrightarrow\,
  2^{3/4}\,\pi^{3/2}\,(\La{+}1)\,\frac{\delta_{C_\gL}}{|C_\gL|}
  = 2^{3/4}\,\pi^{3/2}\,(\La{+}1)\,\delta_\gL^\T \,.
  \labl{a0}
The volume of the conjugacy class is $|C_\gL|\eq 4\pi\kg 
\sin^2\!\psi_\La$, and hence in particular decreases as $\kg^{-1}$
at large level.

In agreement with the general remarks above, \erf{a0} means that for any
$\La$ the brane $\bZ\Lax$ approaches the D0-brane located at $g\eq{\bf1}$.
To see in more detail how this happens, we need to  have a 
closer look at formula \erf{bZ2}. Since $\bZ\Lax$ only depends on the 
coordinate $\psi$, it is natural to regard it as a function on the maximal 
torus, parametrized (modulo the Weyl group) by the angle $\psi\iN[0,\pi]$, 
rather than on \SUtwo. 
When expressing $\bZ\Lax$ as a function of $\psi$, we must take into
account the (Weyl) measure on the maximal torus, which is nothing but the
volume of $C_g$; thus when we want to
visualize the way the brane tends to a delta function, we should study
  \be 
  \bzl(\psi) := |C_g|\,\bZ{\Lax}(\psi) = 4\pi\kg\sin^2\!\psi\,\bZ{\Lax}(\psi)
  \ee
rather than $\bzl$ itself.
Also, when we want to account for the growing radius
$r\eq\sqrt\kg$ of the group manifold $S^3$ (and hence of the torus),
we must measure distances as seen by a `comoving' observer on the group,
i.e.\ use the scaled variable $\sqrt\kg\,\psi$. However, as we are particularly
interested in the vicinity of $\psi\eq0$, 
it is indeed convenient to introduce in addition the `blow-up' coordinate 
$a\,{:=}\,(\kg{+}2)\psi/\pi\eq\psi/\psi_0$. In terms of this parameter we have
  \begin{eqnarray}&
  \bzl(\psi) \big|^{}_{\psi=a\psi_0}
  = \Frac{4\pi\kg}{\sqrt {2\pi^2\kg^{3/2}}}
  \;{^{4}\!\!\!\sqrt{\Frac{2}{\kg+2}}}\,\sqrt{\Frac{\kg{+}2}\pi}\, \sin(a\psi_0)
  {\dsty\sum_{\La'=0}^\kg} \sqrt{ \Frac{\psi_{\La'}}{\sin\psi_{\La'}} }\,
  \sin((\La{+}1)\psi_{\La'})\, \sin(a\psi_{\La'}) \,.
  &\nonumber\\[-.3em]~\label{a7}
  \end{eqnarray}
In the limit $\kg\,{\to}\,\infty$ the $\La'$-summation turns into 
($\kg/\pi$ times) an integration over $t\eq\psi_{\La'}$:
  \be
  \lim_{\kg\to\infty} \bzl(a) = 2^{7/4}\, \sqrt{ \Frac \kg \pi }\,
  a \int_{\!0}^\pi\!\! f_\La(a;t)\, {\rm d}t
  \labl{a8}
with 
  \be
  f_\La(a;t) := \sqrt{\Frac t{\sin t}}\,\sin((\La{+}1)t)\, \sin(at) \,. 
  \ee
For any $\La$ the integral $F_\La(a)\,{:=}\,\int_0^{\pi}\!\!f_\La(a;t)
\,{\rm d}t$ is a continuous function of $a$ independent of \kg; it has
its center of mass is at $a\eq\La{+}1$, i.e.\ at $\psi\eq\psi_\La$, 
the maximum being located at a slightly ($<.01)$ smaller value of $a$. 
According to \erf{a8}, in terms of the coordinate $a$, asymptotically at large 
level the shape grows with the level {\em uniformly\/} as $\sqrt{\kg}$. Indeed, 
the shape stabilizes already at small level; this is illustrated in figure 
\ref{fig:shape} for $\La\eq0$, i.e.\ for the brane closest to the exceptional
conjugacy class $\{{\bf1}\}$, and for $\La\eq5$.
   
In terms of the parameter $\sqrt\kg\,\psi\eq a{/}\sqrt\kg$ which accounts for 
the growing radius, the width of the peak of the function $F_\La$ shrinks with 
the level as $\sqrt\kg$, and so does the distance from its peak to the origin,
as well as the distance between the peaks for any two different branes;
the area under the peak stabilizes for large \kg. Since the distance between 
the peaks for different $\La$, and between peak and origin, shrinks exactly 
at the same rate as the width of the peaks,
even at arbitrarily large level we can distinguish the individual branes
and distinguish their location from the origin. It is in this sense that
in the large limit the branes keep being {\em well separated\/} from each
other, and also well separated from the exceptional conjugacy class.
To put it more sloppily, even at arbitrarily large level the branes insist
on being well located at $\psi_\La\,{\simeq}\,\La/\kg$ rather than at zero.

\vfill

\clearpage 

Obviously these arguments generalize from \SUtwo\ to other groups. In particular
in the limit the summation over weights again reduces to a Riemann integral,
which up to an over-all factor is a level-independent smooth function of 
(appropriately scaled) coordinates on the maximal torus.
Technically the analysis is, however, quite a bit more involved.

\vfill

\begin{figure}[h] 
\begin{center}
\begin{picture}(300,400)
    \put(0,300)  {\scalebox{.30}{\includegraphics{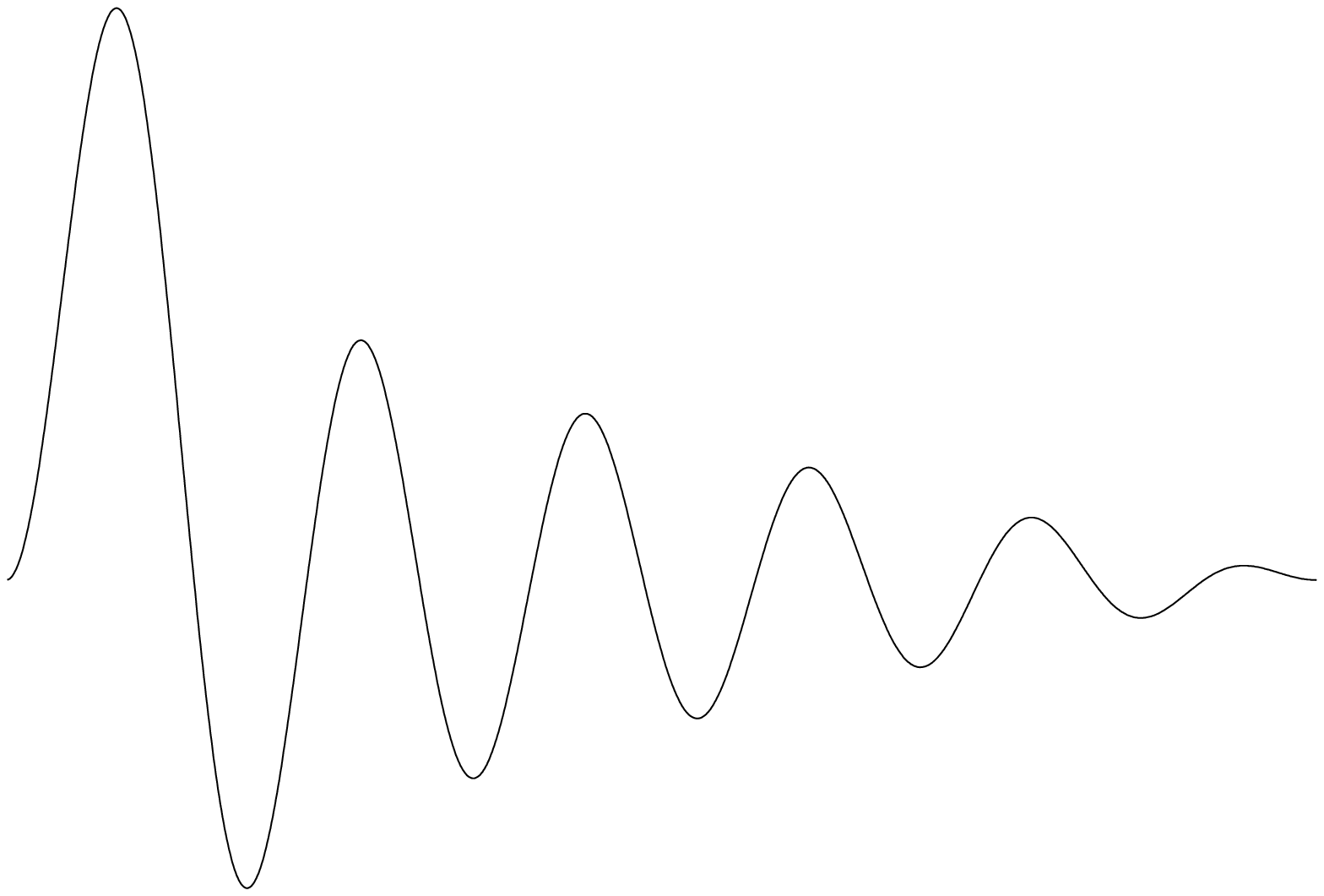}}}
    \put(-2,295) {\begin{picture}(0,0)
        \put(0,0)     {\scalebox{.266}{\includegraphics{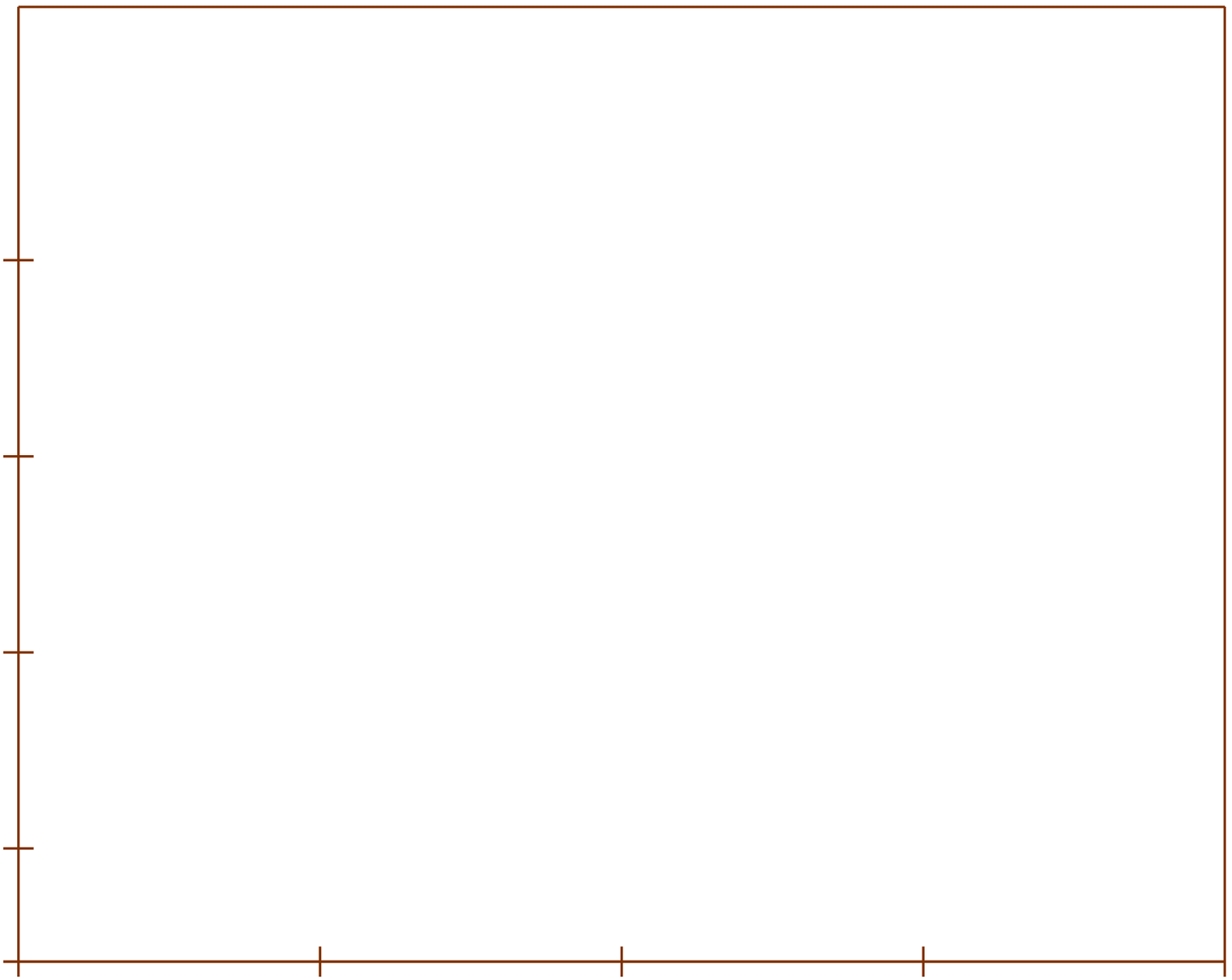}}} 
        \put(-15.5,12){\scriptsize$-25$}
        \put(-4.5,34) {\scriptsize$0$}
        \put(-9,56)   {\scriptsize$25$}
        \put(-9,78)   {\scriptsize$50$}
        \put(0,-6)    {\scriptsize$0$}
        \put(33.5,-7) {\scriptsize$3$}
        \put(67,-7)   {\scriptsize$6$}
        \put(100.5,-7){\scriptsize$9$}
        \put(132,-7)  {\scriptsize$12$}
        \put(99,90)   {\boxx1{10}}
                 \end{picture}}
    \put(190,304.6){\scalebox{.30}{\includegraphics{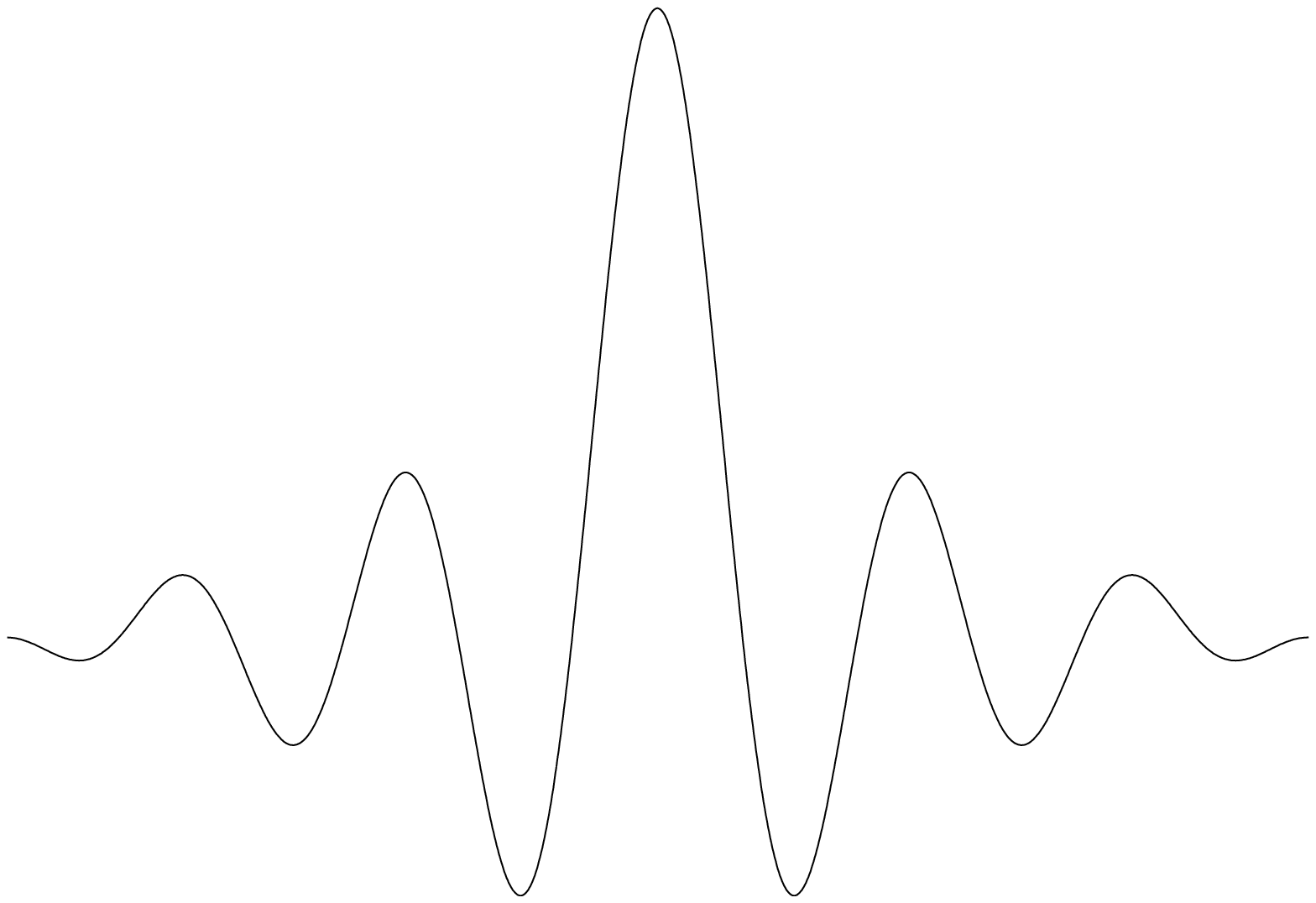}}}
    \put(188,295){\begin{picture}(0,0)
        \put(0,0)     {\scalebox{.266}{\includegraphics{ABaxes.eps}}} 
        \put(-20,12)  {\scriptsize$-100$}
        \put(-5,34)   {\scriptsize$0$}
        \put(-13.5,56){\scriptsize$100$}
        \put(-13.5,78){\scriptsize$200$}
        \put(0,-6)    {\scriptsize$0$}
        \put(33.5,-7) {\scriptsize$3$}
        \put(67,-7)   {\scriptsize$6$}
        \put(100.5,-7){\scriptsize$9$}
        \put(132,-7)  {\scriptsize$12$}
        \put(99,90)   {\boxx5{10}}
                 \end{picture}}
    \put(0,149)  {\scalebox{.35}{\includegraphics{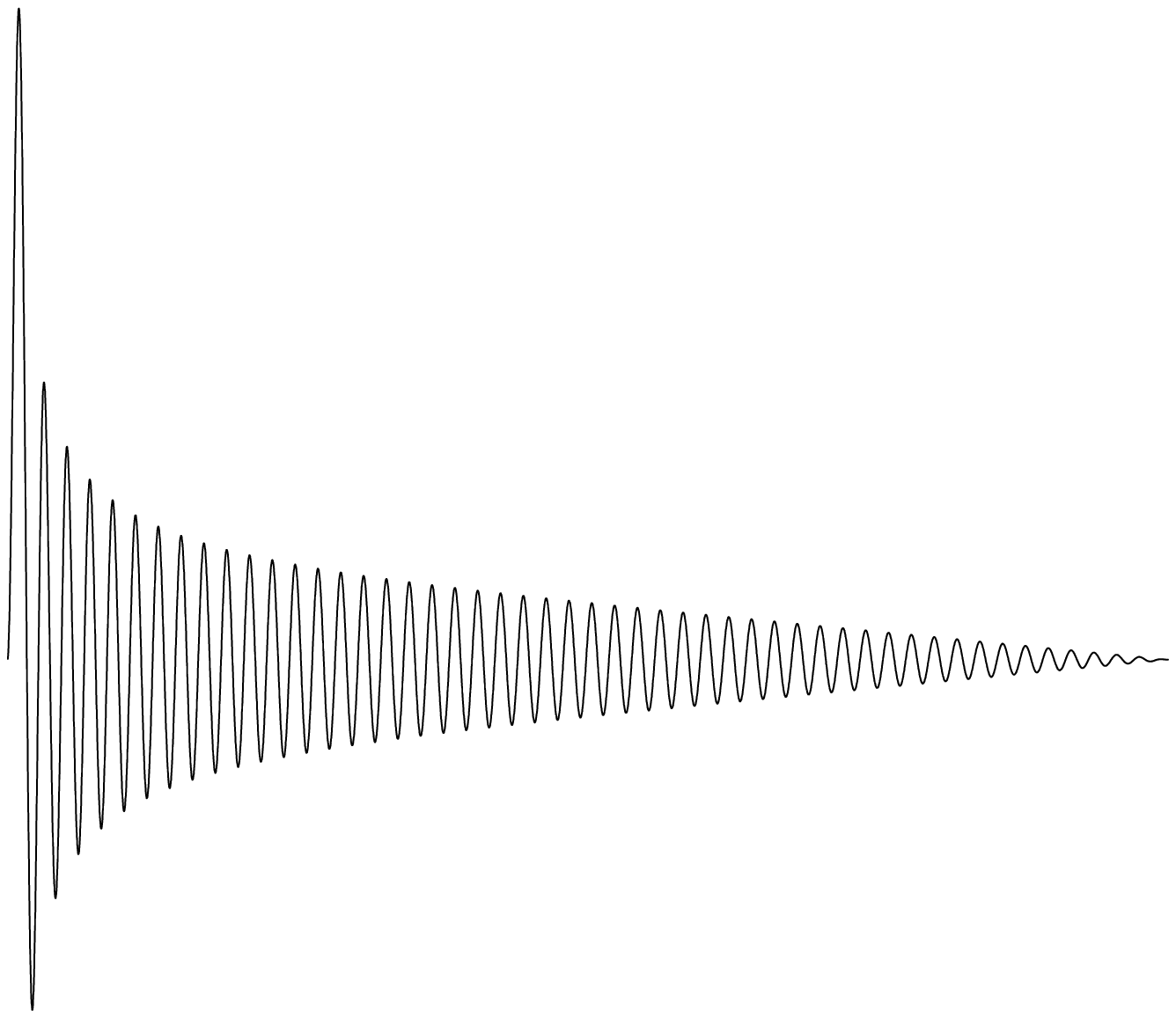}}}
    \put(-2,145) {\begin{picture}(0,0)
        \put(0,0)     {\scalebox{.266}{\includegraphics{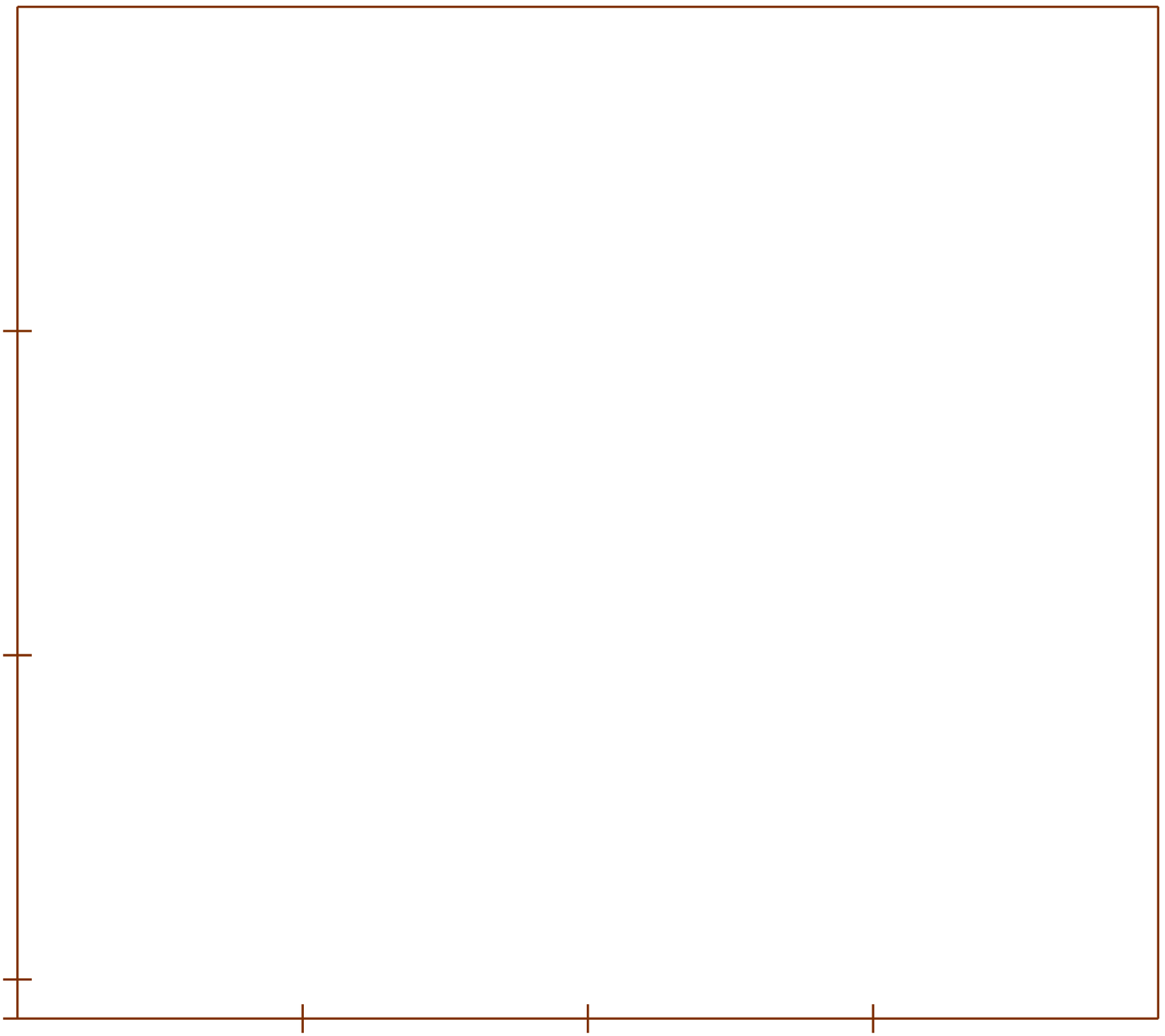}}} 
        \put(-20.5,4) {\scriptsize$-250$}
        \put(-5,42.5) {\scriptsize$0$}
        \put(-13.5,81){\scriptsize$250$}
        \put(0,-6)    {\scriptsize$0$}
        \put(31.5,-7) {\scriptsize$25$}
        \put(65,-7)   {\scriptsize$50$}
        \put(98.5,-7) {\scriptsize$75$}
        \put(130,-7)  {\scriptsize$102$}
        \put(95,103)  {\boxx1{100}}
                 \end{picture}}
    \put(190,156.8){\scalebox{.35}{\includegraphics{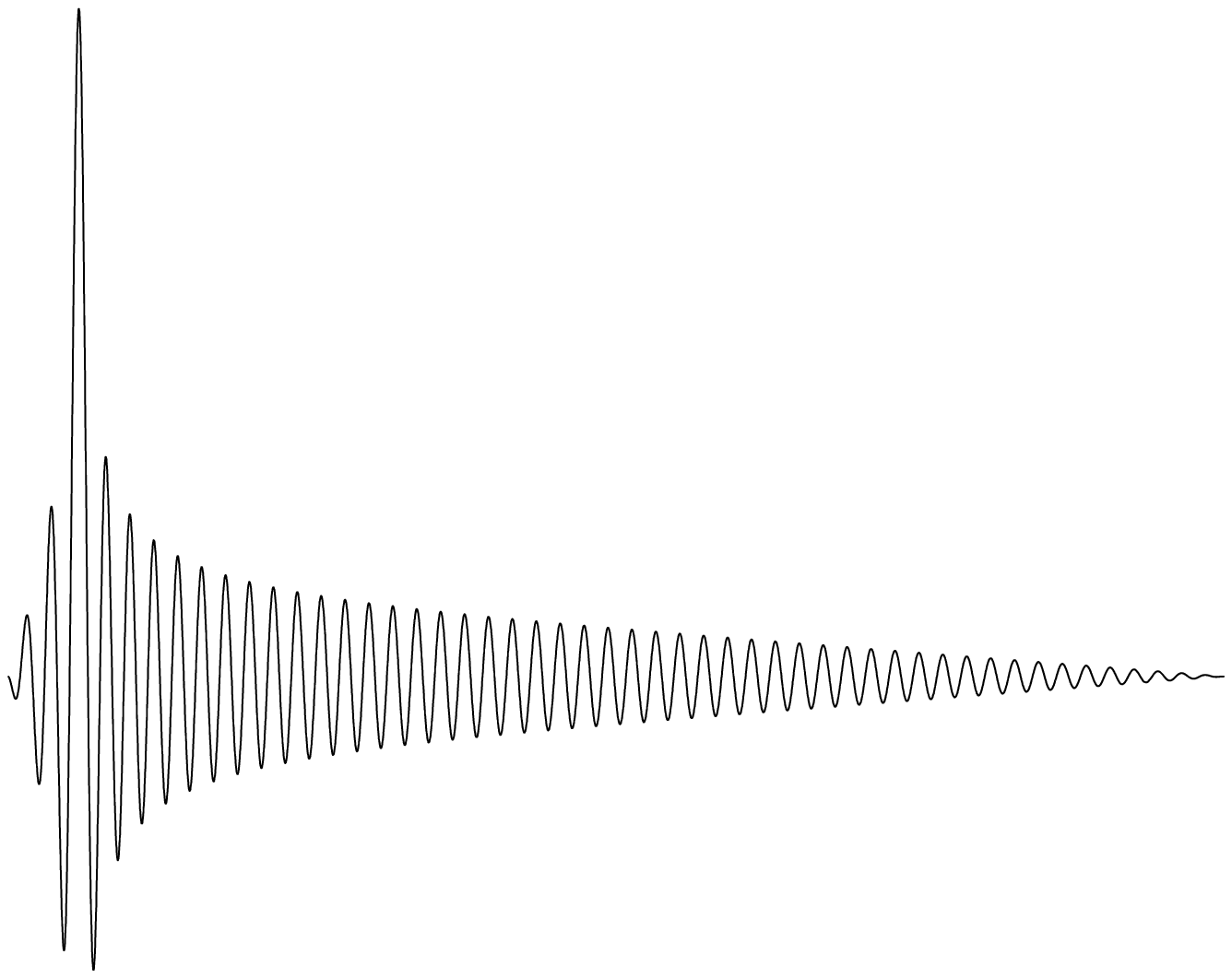}}}
    \put(188,145) {\begin{picture}(0,0)
        \put(0,0)     {\scalebox{.266}{\includegraphics{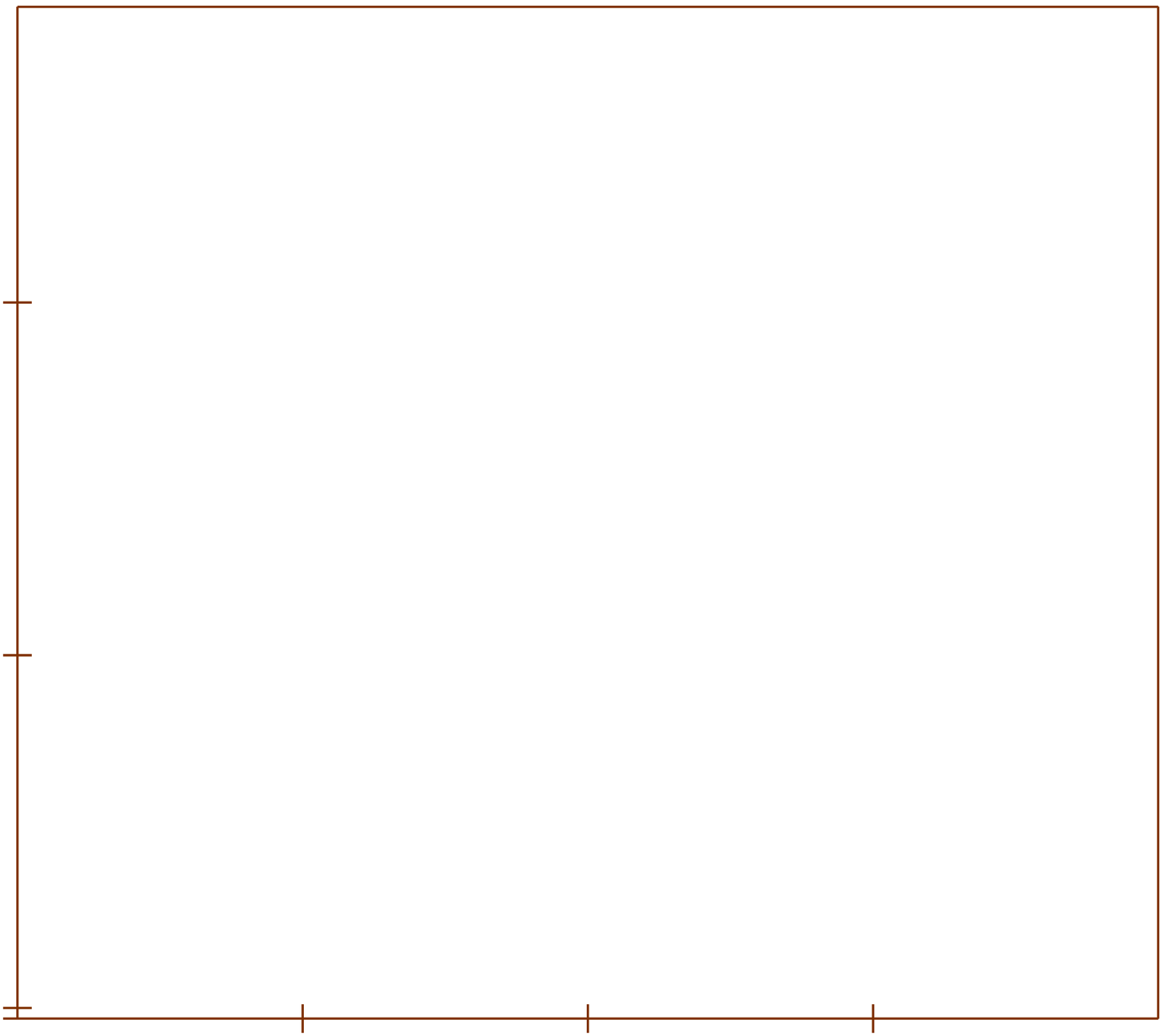}}} 
        \put(-24.5,1) {\scriptsize$-2000$}
        \put(-5,42.5) {\scriptsize$0$}
        \put(-17.5,84){\scriptsize$2000$}
        \put(0,-6)    {\scriptsize$0$}
        \put(31.5,-7) {\scriptsize$25$}
        \put(65,-7)   {\scriptsize$50$}
        \put(98.5,-7) {\scriptsize$75$}
        \put(130,-7)  {\scriptsize$102$}
        \put(95,103)  {\boxx5{100}}
                 \end{picture}}
    \put(0,7)    {\scalebox{.30}{\includegraphics{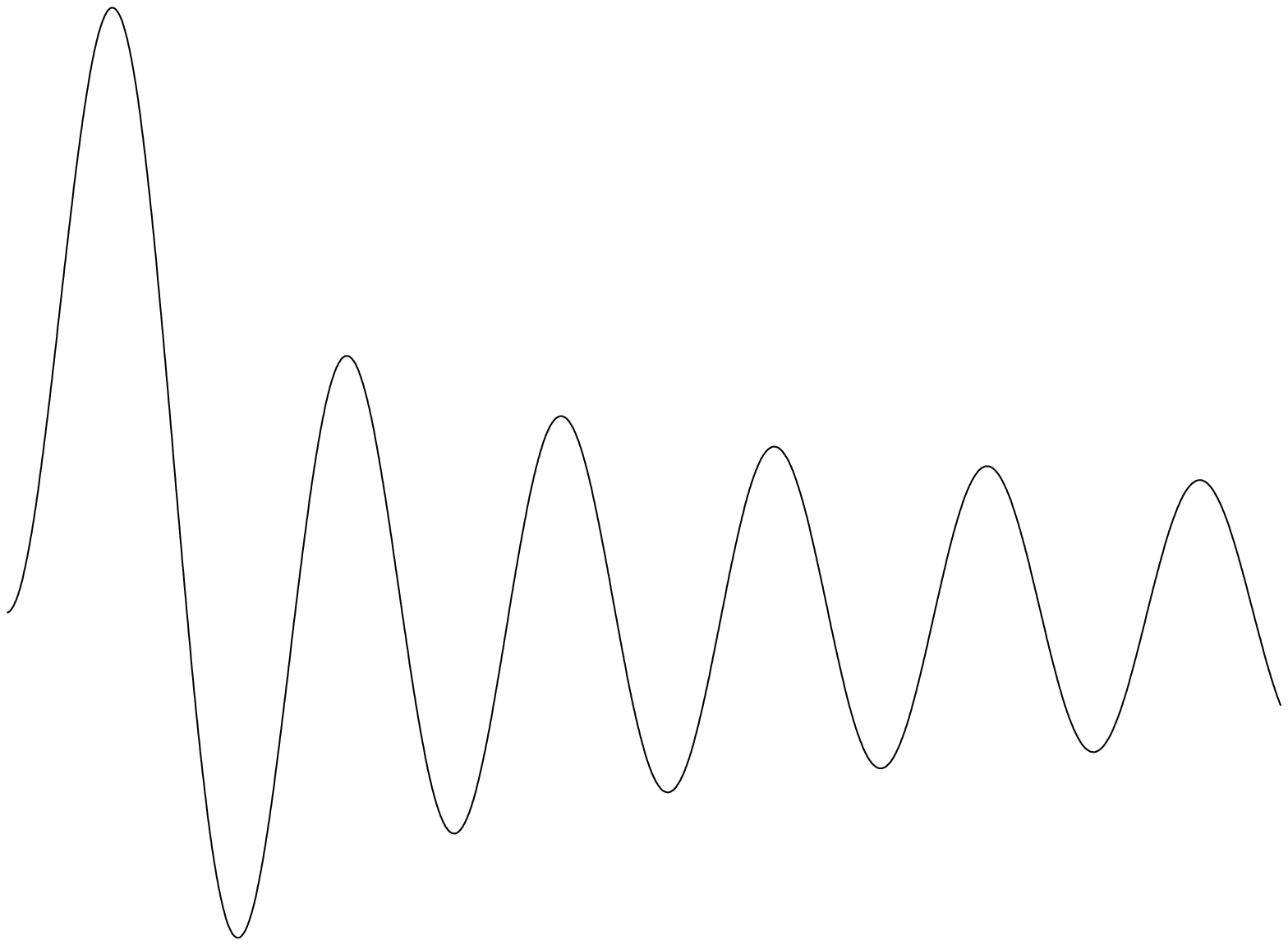}}}
    \put(-2,5)   {\begin{picture}(0,0)
        \put(0,0)     {\scalebox{.266}{\includegraphics{ABaxes.eps}}} 
        \put(-11,12)  {\scriptsize$-1$}
        \put(-4.5,34) {\scriptsize$0$}
        \put(-4.1,56) {\scriptsize$1$}
        \put(-5,78)   {\scriptsize$2$}
        \put(-4,110)  {\scriptsize${\times}10^4$}
        \put(0,-6)    {\scriptsize$0$}
        \put(33.5,-7) {\scriptsize$3$}
        \put(67,-7)   {\scriptsize$6$}
        \put(100.5,-7){\scriptsize$9$}
        \put(132,-7)  {\scriptsize$12$}
        \put(84,90)   {\boxx1{10\,000}}
                 \end{picture}}
    \put(190,14.5){\scalebox{.30}{\includegraphics{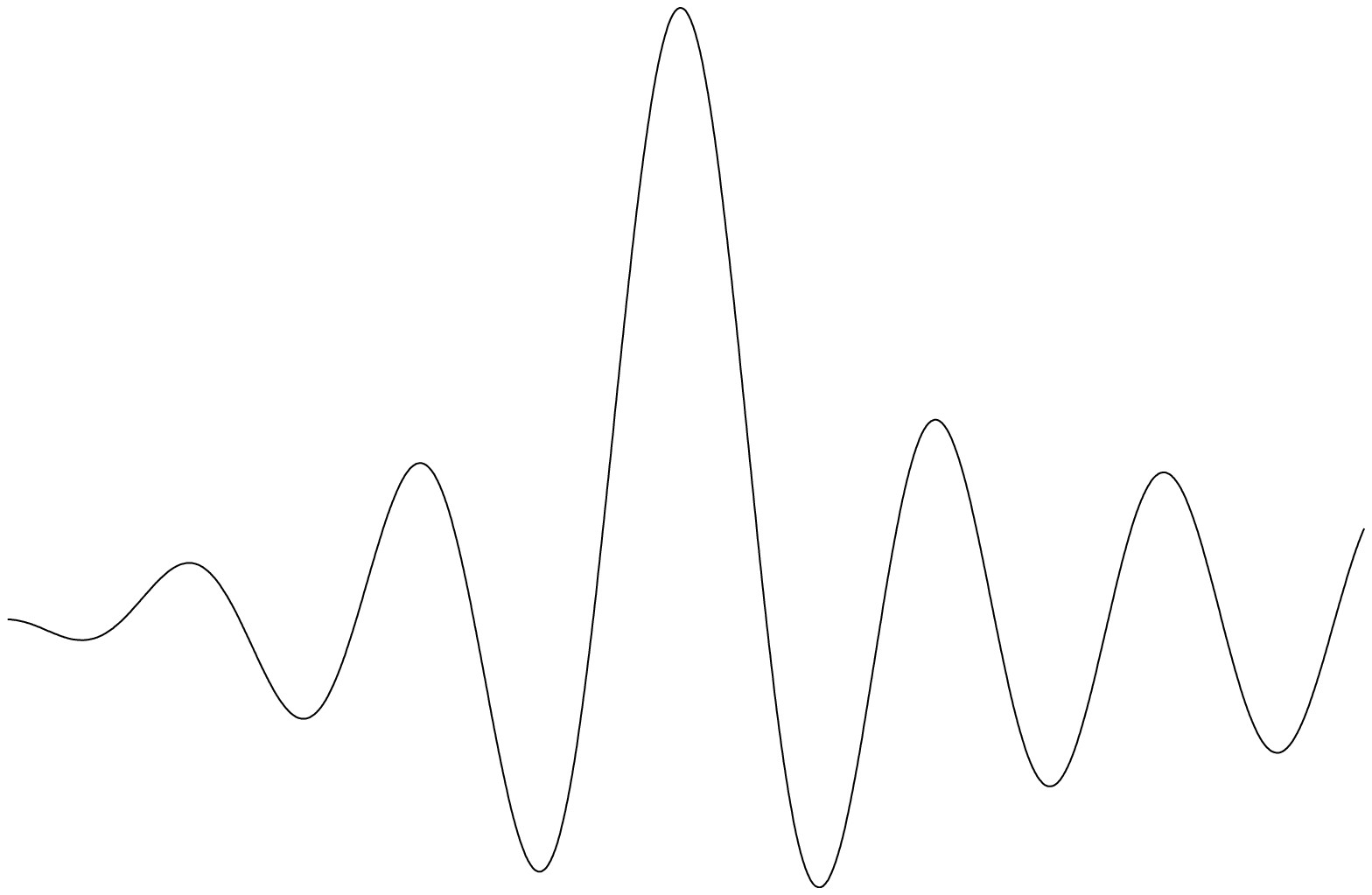}}}
    \put(188,5){\begin{picture}(0,0)
        \put(0,0)     {\scalebox{.266}{\includegraphics{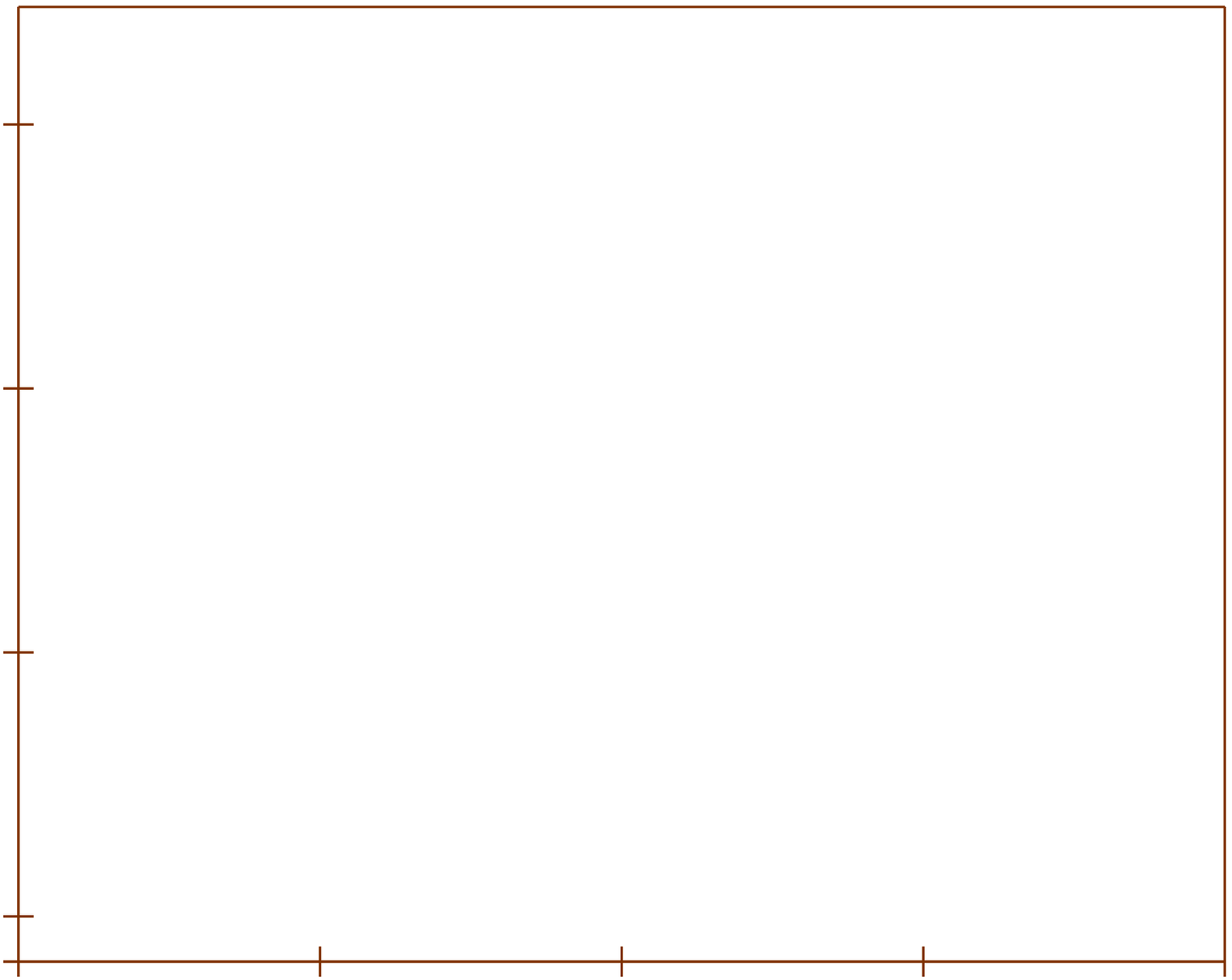}}} 
        \put(-11.5,4.5) {\scriptsize$-1$}
        \put(-4.5,34) {\scriptsize$0$}
        \put(-4.6,63.5) {\scriptsize$1$}
        \put(-5.3,93) {\scriptsize$2$}
        \put(-4,110)  {\scriptsize${\times}10^5$}
        \put(0,-6)    {\scriptsize$0$}
        \put(33.5,-7) {\scriptsize$3$}
        \put(67,-7)   {\scriptsize$6$}
        \put(100.5,-7){\scriptsize$9$}
        \put(132,-7)  {\scriptsize$12$}
        \put(84,90)   {\boxx5{10\,000}}
                 \end{picture}}
\end{picture}
\caption{The brane shape \erf{a7} as a function of the blow-up variable $a$ 
for $\La\eq0$ and $\La\eq5$ at levels $\kg\eq10$, $100$ and $10\,000$.
For $\kg\eq10$ and $\kg\eq100$ the whole range $[0,\kg{+}2]$ of $a$ is shown; 
for $\kg\eq10\,000$ the range $[0,12]$ is displayed instead, in order to 
facilitate comparison with the diagrams in the first row.} 
\label{fig:shape}
\end{center} 
\end{figure}

\vskip.7em
\noindent{\bf Acknowledgements.}\\ We are indebted to Christoph Schweigert for
helpful comments on the manuscript.
\\~

\small

\newcommand\wb{\,\linebreak[0]} \def\wB {$\,$\wb}
 \newcommand\Bi       {\bibitem}
 \renewcommand\J[5]   {{\sl #5\/}, {#1} {#2} ({#3}) {#4} }
 \newcommand\K[6]     {{\sl #6\/}, {#1} {#2} ({#3}) {#4} }
 \newcommand\PhD[2]   {{\sl #2\/}, Ph.D.\ thesis (#1)}
 \newcommand\Mast[2]  {{\sl #2\/}, Master's thesis (#1)}
 \newcommand\Prep[2]  {{\sl #2\/}, pre\-print {#1}}
 \newcommand\BOOK[4]  {{\sl #1\/} ({#2}, {#3} {#4})}
 \newcommand\inBO[7]  {{\sl #7\/}, in:\ {\sl #1}, {#2}\ ({#3}, {#4} {#5}), p.\ {#6}}
 \newcommand\iNBO[7]  {{\sl #7\/}, in:\ {\sl #1} ({#3}, {#4} {#5}) }
 \newcommand\Erra[3]  {\,[{\em ibid.}\ {#1} ({#2}) {#3}, {\em Erratum}]}
 \def\jf    {J.\ Fuchs}
 \def\anip  {Ann.\wb Inst.\wB Poin\-car\'e}
 \def\comp  {Com\-mun.\wb Math.\wb Phys.}
 \def\foph  {Fortschritte\wB d.\wb Phys.}
 \def\ijmp  {Int.\wb J.\wb Mod.\wb Phys.\ A}
 \def\jgap  {J.\wb Geom.\wB and\wB Phys.}
 \def\jhep  {J.\wb High\wB Energy\wB Phys.}
 \def\mama  {Manuscripta\wB math.}
 \def\nupb  {Nucl.\wb Phys.\ B}
 \def\phlb  {Phys.\wb Lett.\ B}
 \def\phrd  {Phys.\wb Rev.\ D}
 \def\phrl  {Phys.\wb Rev.\wb Lett.}
 \def\phrp  {Phys.\wb Rep.}
 \def\ptps  {Progr.\wb Theor.\wb Phys.\wb Suppl.}
 \def\AMS    {{American Mathematical Society}}
 \def\NY     {{New York}}
 \def\PR     {{Providence}}
 \def\SV     {{Sprin\-ger Ver\-lag}}
 \newcommand\qfsm[2] {\inBO{Quantum Fields and Strings: A Course for Mathematicians}
            {P.\ Deligne et al., eds.} \AMS\PR{1999} {{#1}}{{#2}} }

\end{document}